\documentclass[journal]{IEEEtran}
\usepackage{graphicx}
\usepackage{amssymb}
\usepackage{amsmath}
\usepackage{amsthm}
\usepackage[ruled]{algorithm2e}
\usepackage{multirow}
\usepackage{tabu}
\usepackage[numbers,sort&compress]{natbib}
\usepackage{color}

\theoremstyle{plain}

\theoremstyle{definition}
\newtheorem{myDef}{Definition} 

\theoremstyle{remark}

\newcommand{\norm}[1]{\left \Vert #1 \right \Vert}
\newcommand{\abs}[1]{\left \vert #1 \right \vert}
\newcommand{\B}[1]{\mathbb{#1}}
\newcommand{\C}[1]{\mathcal{#1}}

\begin{document}

\title{Robust Platoon Control in Mixed Traffic Flow Based on Tube Model Predictive Control}

\author{Shuo~Feng,
        Ziyou~Song,
        Zhaojian~Li,
        Yi~Zhang,~\IEEEmembership{Member,~IEEE}
        and~Li~Li,~\IEEEmembership{Fellow,~IEEE}% <-this % stops a space

\thanks{This work was supported in part by the National Key Research and Development Program of China under Grant 2018YFB1600600, the National Natural Science Foundation of China (61790565), and Collaboration between China \& Sweden regarding research (2018YFE0102800). \emph{(Corresponding author: Li Li)}}
\thanks{S. Feng is with the Department of Civil and Environmental Engineering, University of Michigan, Ann Arbor, United States (e-mail: fshuo@umich.edu).}
\thanks{Z. Song is with the Department of Electrical Engineering and Computer Science, University of Michigan, Ann Arbor, United States. (e-mail: ziyou@umich.edu).}
\thanks{Z. Li is with the Department of Mechanical Engineering,
	Michigan State University, United States. (e-mail: lizhaoj1@egr.msu.edu).}
\thanks{Y. Zhang and L. Li are with the Department of Automation, Beijing National Research Center for Information Science and Technology (BNRist), Tsinghua University, Beijing 100084, China (e-mail: zhyi@mail.tsinghua.edu.cn; li-li@mail.tsinghua.edu.cn).}
}

%
%\markboth{IEEE Transactions on Vehicular Technology,~Vol.~, No.~, ~2020}%

\maketitle

\begin{abstract}
The design of cooperative adaptive cruise control is critical in mixed traffic flow, where connected and automated vehicles (CAVs) and human-driven vehicles (HDVs) coexist. Compared with pure CAVs, the major challenge is how to handle the prediction uncertainty of HDVs, which can cause significant state deviation of CAVs from planned trajectories. In most existing studies, model predictive control (MPC) is utilized to replan CAVs' trajectories to mitigate the deviation at each time step. However, as the replan process is usually conducted by solving an optimization problem with information through inter-vehicular communication, MPC methods suffer from heavy computational and communicational burdens. To address this limitation, a robust platoon control framework is proposed based on tube MPC in this paper. The prediction uncertainty is dynamically mitigated by the feedback control and restricted inside a set with a high probability. When the uncertainty exceeds the set or additional external disturbance emerges, the feedforward control is triggered to plan a ``tube'' (a sequence of the set), which can bound CAVs' actual trajectories. As the replan process is usually not required, the proposed method is much more efficient regarding computation and communication, compared with the MPC method. {Comprehensive simulations are provided to validate the effectiveness of the proposed framework.}
\end{abstract}


\IEEEpeerreviewmaketitle

\section{Introduction}

\IEEEPARstart{C}{ooperative} Adaptive Cruise Control (CACC) is one of the promising intelligent transportation technologies that contribute to improving traffic flow stability, throughput, and safety \cite{van2006impact, li2014survey, feng2015real, li2015overview, gong2016constrained, zheng2018traffic, sun2019behaviorally, wang2019cooperative}.  In the traffic environment, heterogeneous external disturbances usually exist, such as speed guidance from intersection control \cite{feng2018spatiotemporal, yu2018integrated, yang2019eco} and maneuvers of neighborhood vehicles (e.g., cut-in, cut-out, and merge), which force vehicles to adjust their speed trajectories. Through vehicle-to-vehicle (V2V) and vehicle-to-infrastructure (V2I) wireless communication, CACC can utilize more information to better response to these external disturbances.

Most existing CACC studies focus on pure connected and automated vehicle (CAV) flow \cite{eskandarianresearch}. Given that CAVs and human-driven vehicles (HDVs) will coexist in the future traffic flow for a long period, however, it is critical to design CACC in the mixed traffic flow. To address this issue, several CACC methods in mixed traffic flow have recently been proposed. The basic idea behind these methods is to plan CAVs' trajectories based on the prediction of HDVs. Although various methods can be used for the prediction, such as model-based methods \cite{chen2010markov, jin2018connected, gong2018cooperative} and model-free methods \cite{wang2018capturing, feng2018better, wang2019long}, HDVs usually do not follow deterministic behavioral models and therefore cannot be exactly predicted.

The major challenge of CACC in mixed traffic flow is how to handle the prediction uncertainty of HDVs. Although the uncertainty is usually assumed bounded during one time step, it can accumulate with the increase of prediction time horizon. If CAVs are controlled without adjustment for the accumulated uncertainty, it can cause significant deviation of CAV states (e.g., relative distance and speed) from their planned state trajectories, and therefore violate system constraints such as safety, stability, and string stability \cite{feng2019string, studli2017vehicular, gunter2019model}. 

In most existing studies, model predictive control (MPC) is utilized to replan CAVs' trajectories at each time step (e.g., 0.1 second) \cite{mayne2000constrained, li2017cooperative, gong2018cooperative, chen2018robust}. As the prediction uncertainty is assumed bounded for one time step, MPC methods can implicitly restrict the state deviation of CAVs within a reasonable bound. However, as the replan process is usually conducted by solving an optimization problem with information through inter-vehicular communication, MPC methods suffer from heavy computational and communicational burdens. The communication burden will further damage the CACC performance when time delay and packet loss exist \cite{guo2016communication, wen2018cooperative}.

To address this limitation, this paper proposes a robust platoon control framework based on tube MPC \cite{mayne2001robustifying}. At each time step, the feedback control is utilized to mitigate the deviation of CAV states caused by the uncertainty of HDVs. As no communication is required by the feedback control and its computational cost is negligible, it can utilize a much smaller time step than MPC. Assuming the uncertainty is bounded for one time step as most existing studies do, the state tracking error is proved bounded inside the set (minimal robust positively invariant (mRPI) set) by the feedback control. When external disturbance emerges, instead of optimizing a sequence of CAV states as MPC does, a sequence of mRPI set (i.e., a tube) is determined by solving an optimization problem. The actual state trajectories, which are combined with planned state trajectories and state deviation, are proved bounded by the tube regardless of the actual uncertainty. Therefore, no replan process is required unless additional external disturbance emerges, which significantly reduces the burdens of computation and inter-vehicular communication compared with MPC methods. Since the tube is determined satisfying all constraints, safety, stability, and string stability are guaranteed if the controller is feasible.

\textcolor{black}{
To reduce the method conservativeness and improve the method feasibility, the probabilistic uncertainty bound of HDVs is proposed, and the probabilistic mRPI set is calculated correspondingly. To handle the probabilistic uncertainty bound as well as multiple external disturbances, the event-triggered mechanism is designed for the feedforward control. Consequently, the proposed framework can restrict the uncertainty by the feedback control inside the mRPI set for most of the time (with the probability), and, for the other time when the designed event happens, the feedforward control is triggered. The higher the probability that the uncertainty is inside the bound, the less the feedforward control is triggered, while the higher conservativeness the method could be. Therefore, the framework has the flexibility to balance between the feedback control and feedforward control. 
}

Specifically, the feedback control is designed by the discrete linear quadratic regulator method \cite{bender1987linear}, and the mRPI set is determined by the $\epsilon$-approximation method \cite{rakovic2005invariant}. The feasible set is determined by all constraints, and a tight set is obtained by shrinking the feasible set subtracting the mRPI set. Then, the planning process (i.e., feedforward control) is designed to satisfy the tight set. Simulations are designed to further validate the performance. Results show that the proposed method can guarantee safety, stability, and string stability, and is much more efficient regarding computation and communication, compared with the MPC method.

\textcolor{black}{In summary, the major contribution of this paper is proposing a new framework for cooperative platoon control in mixed traffic flow. Compared with state-of-the-art methods \cite{li2017cooperative, gong2018cooperative} based on MPC, our framework has the following advantages. First, our framework can explicitly handle the prediction uncertainty of human drivers and thus enhance the MPC from an optimization method to a robust optimization method. The robustness can guarantee system constraints such as safety and stability for every external disturbance. Second, our framework is flexible to balance the feedforward control and feedback control. It can gracefully degrade as the pure feedback control if no trajectory plan is available, whereas degrade as the pure feedforward control if the feedback control is too conservative to be feasible. Third, our framework can reduce the trigger frequency of the feedforward control, and thus significantly reduce the burdens of computation and intravehicular wireless communication. It provides huge potentials for designing more effective V2V communication systems and computational resource allocation systems \cite{zheng2015smdp, zhou2019computation, ye2019deep}. }

\textcolor{black}{
The rest of this paper is organized as follows.  The problem is formulated in Section II.  In Section III, the framework of the tube model predictive control is proposed for one external disturbance. The robust platoon control framework is proposed in Section IV including the probabilistic uncertainty bound and the event-triggered mechanism. Performances of the proposed framework are analyzed in Section V. In Section VI, numerical experiments are conducted to validate the effectiveness of the proposed framework. Finally, the paper is concluded in Section VII. 
}

\section{Problem formulation}

\subsection{Notations}
\label{sec_pre}
The field of a real number is denoted by $\B{R}$, whereas $\B{N} = \{1, 2, \dots\}$. For a vector $x \in \B{R}^n$, its $p$-norm is given as
\begin{eqnarray}
&&\norm{x}_p = \left( \sum_{i=1}^n \abs{x_i}^p \right)^{1/p},  p \in [1,\infty) \nonumber\\
&&\norm{x}_{\infty} =  \max_{i} \abs{x_i}.\nonumber
\end{eqnarray}
Given a  Lebesgue measurable signal $x: I \to \B{R}^n$, $\norm{x}^I_{\C{L}_p}$ denotes its $\C{L}_p$ norm defined as
\begin{eqnarray}
&&\norm{x}^I_{\C{L}_p} = \left( \int_I \norm{x(t)}^p_p dt \right)^{1/p} < \infty, p \in [1,\infty) \nonumber \\
&& \norm{x}^I_{\C{L}_\infty} = \sup_{t \in I} \norm{x}_\infty, \nonumber
\end{eqnarray}
where the shorthand notation $\norm{x}_{\C{L}_{\infty}}=\norm{x}^{[0,\infty)}_{\C{L}_{\infty}}$ is used when $I = [0,\infty)$ \cite{zhou1996robust}.
A continuous function $\alpha:[0, a) \to [0, \infty), a\in\B{R^+}$ is said to be of class $\C{K}$ if it is strictly increasing and $\alpha(0) = 0$. We say $x \in \C{L}_{\infty}$ if $\norm{x}_{\C{L}_{\infty}}<\infty$. We recall Minkowski sum for sets $\B{A}, \B{B}$ is $\B{A} \oplus \B{B} = \{ x + y| x \in \B{A}, y\in \B{B}  \}$ and the Pontryagin difference is $\B{A} \ominus \B{B} = \{ x | x + y \in \B{A}, y \in \B{B} \}$. 

\subsection{Scenario Description}

\begin{figure}[h!]
	\centering
	\includegraphics[width=0.45\textwidth]{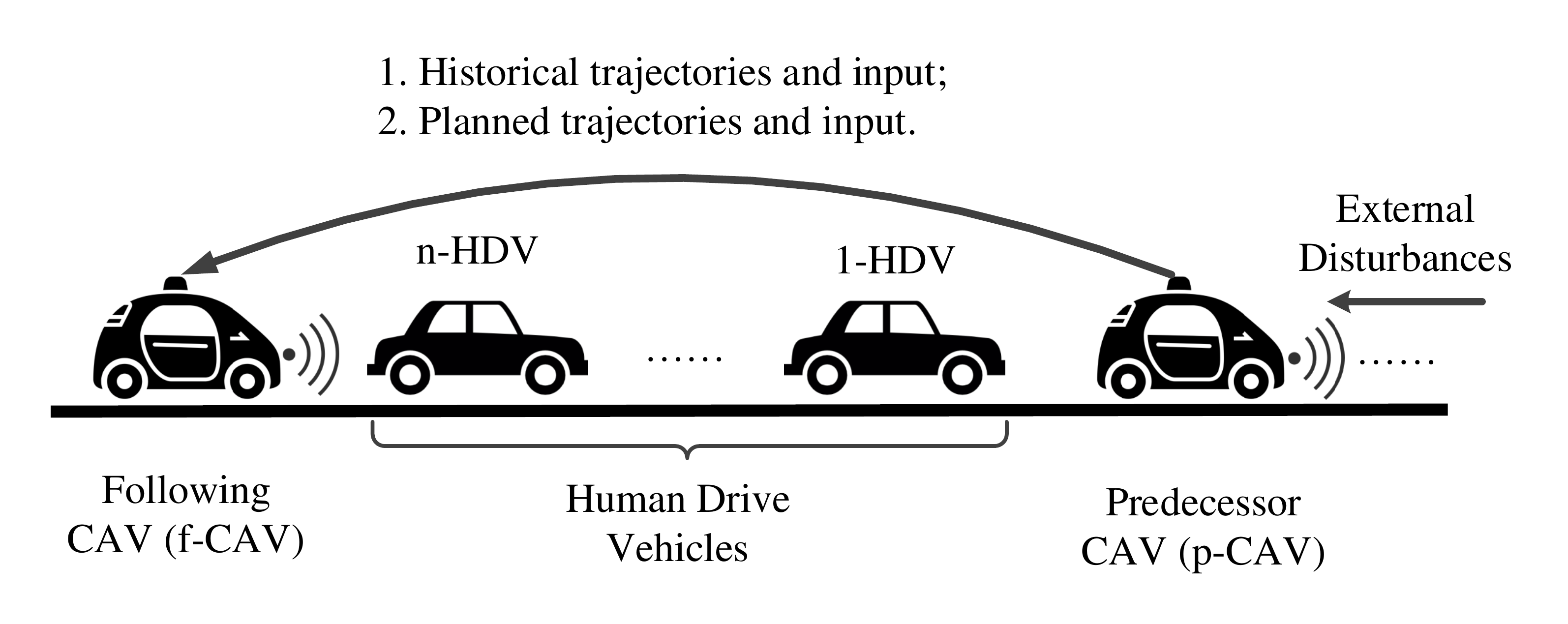}
	\caption{Illustration of a platoon in mixed traffic flow.} 
	\label{fig_platoon}
\end{figure}

As shown in Fig. \ref{fig_platoon}, a sample scenario of mixed traffic flow is studied in this paper, similar to existing studies \cite{gong2016constrained, gong2018cooperative}. The scenario includes a predecessor CAV ($p$-CAV), multiple HDVs, and the following CAV ($f$-CAV). Intervehicular communication exists from $p$-CAV to $f$-CAV. $f$-CAV can measure the distance and speed of its predecessor \textcolor{black}{neighbor} HDV ($n$-HDV) by on-board sensors (e.g., millimeter-wave radars). The situation before $p$-CAV is not specified. There can exist more HDVs before $p$-CAV, so the scenario is a sample of long mixed traffic flow. Without loss of generality, $p$-CAV is assumed as the leading vehicle of a platoon.

When external disturbances emerge, such as speed guidance from intersection control and maneuvers of neighborhood vehicles (e.g., cut-in, cut-out, and merge), CAVs are forced to adjust their trajectories. The trajectory of $p$-CAV can be planned by various methods (e.g., one-step MPC) to eliminate the external disturbances. With the information of historical and planned trajectories of $p$-CAV, how to control $f$-CAV efficiently and robustly remains a huge challenge, which is the focus of this paper.

\subsection{CAV Modeling}
The discrete vehicle dynamic of CAVs can be constructed by the continuous-time dynamics with Zero Order Hold method \cite{ogata1995discrete} \textcolor{black}{and linearization techniques \cite{zheng2016stability}. In this paper, the second-order linear model is applied, which has been widely used before \cite{yanakiev1996simplified, naus2010string, li2015overview, feng2019string}. Denote $s$,  $v$, $u$ as the position, speed, and acceleration.} With a sampling time interval of $\tau$, the discrete dynamic is obtained as
\begin{eqnarray}
\begin{aligned}
\label{eq_dyn_CAV}
x_i(k+1) = A x_i(k) + Bu_i(k), i \in \{p, f\}
\end{aligned}
\end{eqnarray}
where $x_p$ and $x_f$ denote the states of $p$-CAV and $f$-CAV respectively, and
\begin{eqnarray}
x_i = \left[\begin{matrix}
s_i\\v_i
\end{matrix}\right],
A = \left[\begin{matrix}
1 &\tau\\ 
0 &1\\
\end{matrix}\right],
B = \left[\begin{matrix}
0.5\tau^2\\
\tau\\
\end{matrix}\right].\nonumber
\end{eqnarray}
If the planned state and acceleration are denoted as $\bar{x}$ and $\bar{u}$, the dynamic can be denoted as
\begin{eqnarray}
\begin{aligned}
\label{eq_dyn_CAV_plan}
\bar{x}_i(k+1) = A \bar{x}_i(k) + B\bar{u}_i(k), i \in \{p, f\}. 
\end{aligned}
\end{eqnarray}
Subtracting Eq. (\ref{eq_dyn_CAV}) by Eq. (\ref{eq_dyn_CAV_plan}), dynamic of the state deviation can be denoted as
\begin{eqnarray}
\begin{aligned}
\label{eq_dyn_CAV_de}
\tilde{x}_i(k+1) = A \tilde{x}_i(k) + B \tilde{u}_i(k), i \in \{p, f\}, \nonumber
\end{aligned}
\end{eqnarray}
where
\begin{eqnarray}
\begin{aligned}
\tilde{x}_i(k) = x_i(k) - \bar{x}_i(k), \nonumber\\
\tilde{u}_i(k) = {u}_i(k) - \bar{u}_i(k). \nonumber
\end{aligned}
\end{eqnarray}

The tracking error of $f$-CAV is defined as
\begin{eqnarray}
\label{eq_err}
\begin{aligned}
e &= [e_{s},e_{v}]^T,
\end{aligned}
\end{eqnarray}
where $e_s$ denotes the position tracking error, and $e_v$ denotes the speed tracking error. In this paper, constant time headway range policy \cite{chien1992automatic, zhou2005range} is applied, so the tracking error is further defined as
\begin{eqnarray}
\begin{aligned}
&e_{s}(k) = s_{n}(k) - s_f(k) - h \cdot v_f(k), \nonumber\\
&e_{v}(k) = v_{n}(k) - v_f(k),
\end{aligned}
\end{eqnarray}
where $k$ denotes the discrete-time step, $s_n$ and $s_f$ denote the position of the $n$-HDV and $f$-CAV respectively, $v_n$ and $v_f$ denote their speeds, and $h$ denotes the constant time headway. To make the paper concise, the tracking errors are represented compactly as
\begin{eqnarray}
\label{eq_err_M}
\begin{aligned}
e(k) &= x_{n}(k) +C x_f(k), 
\end{aligned}
\end{eqnarray}
where
\begin{eqnarray}
C = \left[\begin{matrix}
-1 &-h\\ 
0 &-1 \\
\end{matrix}\right]. \nonumber
\end{eqnarray}

\subsection{HDV Prediction}
Behaviors of HDVs are usually predicted for optimizing state trajectories of CAVs. Let $\bar{x}_n(k)$ denote the predicted state of $n$-HDV at time step $k$, then the predicted tracking error of $f$-CAV, denoted as $\bar{e}(k)$, can be calculated by 
\begin{eqnarray}
\label{eq_err_determine}
\begin{aligned}
\bar{e}(k) &= \bar{x}_{n}(k) +C \bar{x}_f(k).
\end{aligned}
\end{eqnarray}

Although various methods can be used for the prediction, HDVs usually do not follow deterministic behavioral models, and therefore prediction uncertainty exists. Let $\tilde{x}_n(k)$ denote the prediction uncertainty as
\begin{eqnarray}
\begin{aligned}
\tilde{x}_n(k) &= x_n(k) - \bar{x}_{n}(k). \nonumber
\end{aligned}
\end{eqnarray}
Then the actual tracking error is influenced as
\begin{eqnarray}
\label{eq_err_decompose}
\begin{aligned}
e(k) &= x_n(k) + C x_f(k), \\
&= \bar{x}_n(k)  + C \bar{x}_f(k) + \tilde{x}_n(k) + C \tilde{x}_f(k), \\
&= \bar{e}(k) + \tilde{e}(k),
\end{aligned}
\end{eqnarray}
where $\tilde{e} = \tilde{x}_n + C \tilde{x}_f$ denotes the deviation of the tracking error from its planned value. 

%If the deviation accumulates significantly, performances of CAVs will be hugely damaged.

\subsection{Control Objectives}
In this paper, the objectives of CACC include safety, stability, and string stability, which are elaborated in this subsection.

For a pure CAV flow, one control objective is to ensure all vehicles in the same group to move at a constant speed, while maintaining the desired spaces between adjacent vehicles, i.e., keep the tracking error as zero \cite{horowitz2000control}. To this end, two types of stability have been proposed, i.e., individual stability and string stability. Stability describes vehicles converging to given trajectories, while string stability describes that the disturbances are not amplified along the string of vehicles, which is critical for traffic flow stability \cite{talebpour2016influence, ke2018new, gunter2019model}. Definitions and analysis methods of string stability can be found in review studies  \cite{feng2019string, studli2017vehicular}.

Different from pure CAV flow, not all vehicles can be controlled in mixed traffic flow, so the definitions should be modified correspondingly as follows:
\begin{myDef}
	\label{def_individual}
	(Individual Stability): A mixed platoon is said to be individual stable if 
	\begin{eqnarray}
	\begin{aligned}
	\dot{v}_n(t) = 0,  \forall t \ge 0   \Rightarrow \lim_{t \to \infty} {e}(t) = 0, \nonumber
	\end{aligned}
	\end{eqnarray}
	where $v_n$ denotes the speed of $n$-HDV.
\end{myDef}

\begin{myDef}
	\label{def_lp}
	($\C{L}_p$ String Stability): A mixed platoon is said to be $\C{L}_p$ string stable if there exist class $\C{K}$ function $\alpha$ and constant $c>0 $, such that, for any initial external disturbance of the $p$-CAV satisfying
	\begin{eqnarray}
	\begin{aligned}
	\abs{{e}_{s, p}(0)} <c, \nonumber
	\end{aligned}
	\end{eqnarray}
	the solution ${e}_{s}(t)$ exists for all $t>0$ and satisfies
	\begin{eqnarray}
	\label{eq_def_lp}
	\begin{aligned}
	e \in \B{D}  &= \left \{e \in \B{R}^2: \norm{e_s(t)}_{\C{L}_p} \le \alpha \left( \abs{{e}_{s, p}(0)} \right) \right \},
	\end{aligned}
	\end{eqnarray}
	where ${e}_{s, p}(0)$ denotes the initial position disturbance of the $p$-CAV.
\end{myDef}

To better understand the definitions, we further explain their properties. Definition \ref{def_individual} requires that  tracking errors of CAVs should converge to zero if HDVs keep the speed constant. This property guarantees that the controlled CAVs can converge to the given trajectory if there is no external disturbance. Definition \ref{def_lp} requires that tracking errors of CAVs are bounded if the initial external disturbance is bounded. In existing studies, $\C{L}_2$ and $\C{L}_{\infty}$ norms are usually utilized, which refer to the energy and maximal amplitude respectively.

Besides stability and string stability, safety is another critical control objective. Moreover, the speed and acceleration should be constrained. Therefore, constraints of $f$-CAV are summarized as
\begin{eqnarray}
\begin{aligned}
\label{set_origin}
e \in \B{E}  &= \{e \in \B{R}^2: e \in \B{D}, -d_{min} \le e_{s}, \\
&v_n - v_{max} \le  e_v \le v_n - v_{min} \}, \\
u_f \in \B{U} &= \{ u_f \in \B{R}: -u_{max} \le u_f \le u_{max} \}, 
\end{aligned}
\end{eqnarray}
where $e \in \B{D}$ is the requirement of string stability (see Eq. (\ref{eq_def_lp})), $-d_{min}$ represents the minimal distance for safety requirement, $v_{min}$ and $v_{max}$ denote the minimum and maximum speeds respectively, and $u_{max}$ denotes the maximum acceleration. 

\subsection{Challenge Brought by Prediction Uncertainty}
\label{sec_challenge}
In most existing studies, the prediction uncertainty for one time step is assumed bounded explicitly or implicitly. Adopting this assumption, the dynamic of $\tilde{e}$ is derived as
\begin{eqnarray}
\label{eq_sys_err_ori}
\begin{aligned}
\tilde{e}(k+1) &= \tilde{x}_n(k+1) + C \tilde{x}_f(k+1), \\
&= A \tilde{x}_n(k) + \Delta \tilde{x}_n(k) + C (A \tilde{x}_f(k) + B\tilde{u}_f(k)), \\
&= A \tilde{e}(k) + CB\tilde{u}_f(k) + \Delta \tilde{x}_n(k), 
\end{aligned}
\end{eqnarray}
where $\Delta \tilde{x}_n(k)$ denotes the prediction uncertainty during the time step $(k\tau, (k+1)\tau]$, and it is assumed bounded as
\begin{eqnarray}
\label{eq_dis_bound}
\Delta \tilde{x}_n(k) \in \B{W} = \left\{\B{R}^2: \norm{\Delta \tilde{x}_n(k)}_{\infty} \le c_{\omega}\right\}.
\end{eqnarray}

Indicated by Eq. (\ref{eq_sys_err_ori}), if $f$-CAV is controlled only by its planned acceleration, i.e., $\tilde{u}_f(k) = 0$, the deviation of the tracking error $\tilde{e}$ will accumulate with the time horizon. As a result, the tracking error will also accumulate as shown in Eq. (\ref{eq_err_decompose}), violating the constraints as shown in Eq. (\ref{set_origin}). The robust platoon control is proposed in this paper to address this challenge.

%To address this challenge,  $\tilde{u}_f(k)$ needs to be determined nonzero to mitigate the tracking error. From this perspective, model predictive control (MPC) methods essentially determine $\tilde{u}_f(k)$ by the replanning process at each time step.

\section{Tube Model Predictive Control}
\begin{figure}[h!]
	\centering
	\includegraphics[width=0.5\textwidth]{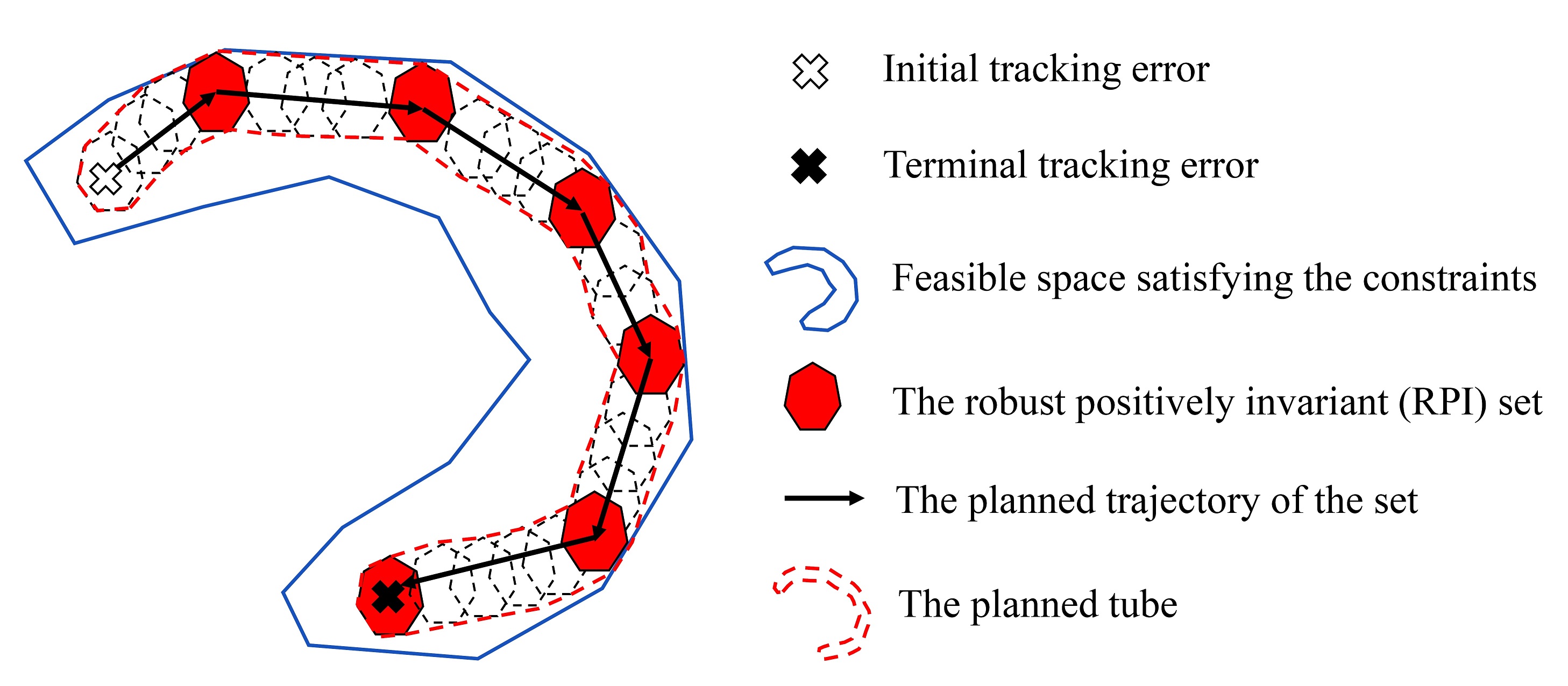}
	\caption{Illustration of the tube method.} 
	\label{fig_TubeMethod}
\end{figure}

In this section, the control framework is proposed for one external disturbance based on tube MPC \cite{mayne2001robustifying, mayne2016robust, feng2019tube} including feedback control and feedforward control. As illustrated in Fig. \ref{fig_TubeMethod}, the feedforward control determines a sequence of set (a tube), while the feedback control dynamically mitigates the deviation ($\tilde{e}$) to restrict the actual tracking error ($e=\bar{e}+\tilde{e}$) inside the tube. Because the tube belongs to the feasible set, all constraints are satisfied by the actual tracking error. The actual acceleration of $n$-CAV is determined by $u_f = \bar{u}_f +\tilde{u}_f$, where $\bar{u}_f$ is planned by the feedforward control, and $\tilde{u}_f$ is determined by the feedback control at each time step. Different from MPC, which replans at each time step, the feedforward control in the proposed method plans $\bar{u}_f$ at the first step when external disturbances emerge, and no replan is required until next external disturbance.

Specifically, the initial tracking error is caused by the external disturbance, and the terminal tracking error is zero according to the stability objective. The feedback control is designed by the discrete  linear quadratic regulator in Subsection \ref{sec_feedback}. The feasible set is determined by all constraints as shown in Eq. (\ref{set_origin}). To reduce the conservativeness of the tube, the minimal robust positively invariant (mRPI) set is defined and estimated by $\epsilon$-approximation method in Subsection \ref{sec_mRPI}. The planned trajectory of the set ($\bar{e}$) is determined by solving an optimization problem with a tight feasible set (i.e., feedforward control) in Subsection \ref{sec_feedforward}.

\subsection{Feedback Control Design}
\label{sec_feedback}
At each time step, the feedback control mitigates the deviation, $\tilde{e}(k)$, by determining $\tilde{u}_f(k)$ as
\begin{eqnarray}
\label{eq_uf}
\tilde{u}_f(k) = K \tilde{e}(k), 
\end{eqnarray}
where $K$ denotes the feedback gain. Substituting Eq. (\ref{eq_uf}) into Eq. (\ref{eq_sys_err_ori}), the dynamic is derived as
\begin{eqnarray}
\label{eq_sys_err}
\tilde{e}(k+1) = A_K \tilde{e}(k) +  \Delta \tilde{x}_n(k), 
\end{eqnarray}
where $A_K = A+CBK$. 

To determine the feedback gain $K$, the discrete linear quadratic regulator problem is formulated and solved as \cite{bender1987linear}:
\begin{eqnarray}
\label{prob_feedback}
\begin{aligned}
\min_{K}  J &= \sum_{0}^{\infty} \left \{ \norm{Q \tilde{e}_{s}(k)}_2  + \norm{L \tilde{e}_{v}(k)}_2 + \norm{R \tilde{u}_f(k)}_2 \right \}, 
\end{aligned}
\end{eqnarray}
where the weighting matrices, i.e., $Q, L$, and $R$, are symmetric and positive definite. 

\subsection{Minimal Robust Positively Invariant Set}
\label{sec_mRPI}
It can be proved that the feedback control in Eq. (\ref{eq_uf}) can bound the deviation $\tilde{e}$ in Eq. (\ref{eq_sys_err}) into a set, denoted as robust positively invariant (RPI) set, which was defined as \cite{blanchini1999set}:
\begin{myDef}
	\label{def_RPI}
	(RPI set): 
	The set $\B{Z} \subset \B{R}^2$ is a robust positively invariant (RPI) set of the system (\ref{eq_sys_err}) if $A_K \tilde{e} + \Delta \tilde{x}_n \in \B{Z}$ for all $\tilde{e} \in \B{Z}$ and all $\Delta \tilde{x}_n\in \B{W}$, i.e., if and only if $A_{K}\B{Z} \oplus \B{W} \subset \B{Z}$.
\end{myDef}

As shown in Fig. \ref{fig_TubeMethod}, the size of the tube is determined by the set. To reduce the conservativeness of the tube, the minimal RPI set is defined as \cite{blanchini1999set}:
\begin{myDef}
	(Minimal RPI set): 
	The mRPI set $\B{Z}$ of the system (\ref{eq_sys_err}) is the RPI set in $\B{R}^2$ that is contained in every closed RPI set of the system (\ref{eq_sys_err}).
\end{myDef}

To determine the mRPI set, it can be equivalently defined as $\B{Z} = \lim_{s \to \infty}F_s$, where
\begin{eqnarray}
\label{eq_Z}
\begin{aligned}
\label{eq_mRPI}
F_s = \bigoplus\limits_{i=0}^{s-1} A_K^i \B{W}, F_0 = \{0\}.
\end{aligned}
\end{eqnarray}
Although it is generally impossible to obtain an explicit characterization of $\B{Z}$ using Eq. (\ref{eq_Z}) \cite{gayek1991survey}, an estimation of $\B{Z}$ can be obtained. In this paper, the $\epsilon$-approximation method is applied to estimate an outer convex set of the mRPI set, i.e., $\B{F} \supset \B{Z}$. By decreasing the difference between $\B{F}$ and $\B{Z}$, the outer convex set can well estimate the mRPI set. The details of the $\epsilon$-approximation method can be found in \cite{rakovic2005invariant}. 

\subsection{Feedforward Control Design}
\label{sec_feedforward}
The goal of the feedforward control is to determine a tube satisfying all constraints as shown in Eq. (\ref{set_origin}) as
\begin{eqnarray}
\begin{aligned}
e \in \B{E}, u_f \in \B{U}. \nonumber
\end{aligned}
\end{eqnarray}
It is essential to determine the planned trajectory, $\bar{e}$, satisfying the tight constraints as
\begin{eqnarray}
\label{eq_tight_set}
\begin{aligned}
\overline{\B{E}} = \B{E} \ominus \B{F}, \overline{\B{U}} = \B{U} \ominus K \B{F}. \nonumber
\end{aligned}
\end{eqnarray}

To determine the trajectory of $\bar{e}$, the optimization problem is formulated as
\begin{eqnarray}
\label{pro_nom}
\begin{aligned}
\min_{\bar{u}_f(1), \cdots, \bar{u}_f(N_p)}  J = \sum_{1}^{N_p} \left\{ \norm{G \bar{e}(k)}_2 +  \norm{F \bar{u}_f(k)}_2 \right \}, 
\end{aligned}
\end{eqnarray}
subject to constraints and dynamic of $\bar{e}$ in Eq. (\ref{eq_err_determine}) as
\begin{eqnarray}
\label{sete_all_nom}
\begin{aligned}
&\bar{e}(k) \in \overline{\B{E}} \cap \B{E}_{t},\\
&\bar{u}_f(k) \in \overline{\B{U}} \cap \B{U}_{t},\\
&\bar{e}(k) = \bar{x}_{n}(k) +C \bar{x}_f(k).
\end{aligned}
\end{eqnarray}
where $k = 1, \cdots, N_p$, $G$ and $F$ are symmetric and positive define weighting matrices, and $\B{E}_{t}, \B{U}_{t}$ are terminal constraints for the stability objective as
\begin{eqnarray}
\label{set_term}
\begin{aligned}
\bar{e} \in \B{E}_t &= \{\bar{e} \in \B{R}^2: \bar{e}(N_p) = 0 \},\\
\bar{u}_f \in \B{U}_{t} &= \{\bar{u}_f \in \B{R}: \bar{u}_f(N_p) = 0\}.
\end{aligned}
\end{eqnarray}
Note $\bar{x}_n$ denotes the prediction of $n$-HDV which relies on the transmitted information from the $p$-CAV.

\textcolor{black}{$N_p$ denotes the predictive steps and should be selected properly as the balance of problem feasibility and computational complexity. A larger $N_p$ indicates more steps that can be used to eliminate the initial error and thus makes the problem in Eq. (\ref{pro_nom}-\ref{set_term}) easier to be solved (higher feasibility). As more steps require to be predicted, however, a larger $N_p$ also has a heavier computational burden. In this paper, a simple yet effective policy is applied for determining $N_p$: an initial value is selected by experience, and, if the problem is infeasible, the value is increased gradually until the problem becomes feasible or the value reaches a threshold ($N$), such as $N_p \leftarrow 2N_p$ if $N_p<N$. The problem is identified as infeasible if it is still infeasible for $N_p=N$.}

\subsection{Algorithm and Limitation}
\label{sec_algorithm}
\textcolor{black}{
	The control framework for one external disturbance can be summarized in Algorithm \ref{alg_tube}. The feedback gain, uncertainty bound, mRPI set, and tight sets are pre-determined, so the computational complexity is negligible. Without external disturbances or no trajectory plan is available ($k>N_p$), the controller gracefully degrades as the pure feedback control, resulting in a car-following behavior.
}

\begin{algorithm}
	\caption{Algorithm for one external disturbance}%算法名字
	\label{alg_tube}
	\LinesNumbered %要求显示行号	
	Compute the feedback gain $K$ by solving the problem in Eq. (\ref{prob_feedback}) as in \cite{bender1987linear}\;
	Determine the uncertainty bound $\B{W}$\;
	Compute the mRPI set $\B{F}$ as in \cite{rakovic2005invariant} and the tight sets $\overline{\B{E}}, \overline{\B{U}}$ by Eq. (\ref{eq_tight_set})\;
	\If{The external disturbance emerges}
	{
		Compute the feedforward control $\bar{u}_f, \bar{e}$ by solving the optimization problem in Eq. (\ref{pro_nom}-\ref{set_term}), where $N_p$ is determined as discussed in Subsection \ref{sec_feedforward}\;
		\For{$k=1$ to $N_p$}
		{
			Observe the actual tracking error $e(k)$\;
			Compute the deviation $\tilde{e}(k)=e(k)-\bar{e}(k)$\;
			Compute the feedback control $\tilde{u}_f(k)= K \cdot  \tilde{e}(k)$\;
			Compute the actual acceleration $u_f(k) = \bar{u}_f(k) + \tilde{u}_f(k)$\;
			Implement the actual acceleration $u_f(k)$\;
		}	
		\textcolor{black}{
		Observe the actual tracking error $e(k)$\;
		Compute the feedback control $\tilde{u}_f(k)= K \cdot  e(k)$\;
		Implement the feedback acceleration $\tilde{u}_f(k)$.
		}		
	}

\end{algorithm}

\textcolor{black}{
	One major limitation is the potential conservativeness of the uncertainty bound $\B{W}$. In Algorithm \ref{alg_tube}, the prediction uncertainty of HDVs for one time step is assumed bounded as in Eq. (\ref{eq_dis_bound}). The assumption is mild as the time interval is usually small. In practice, however, the calculation of the uncertainty bound could be intractable, because it is influenced by types of drivers, and even the same driver can have extreme behaviors. If considering all these possible situations, the bound $\B{W}$ could be so conservative that the optimization problem in Eq. (\ref{pro_nom}-\ref{set_term}) is infeasible for $N_p=N$. We will solve the limitation in the next section.
}

\section{Robust Platoon Control}
\textcolor{black}{
	In this section, the framework of robust platoon control is proposed for multiple external disturbances based on the tube MPC. In Subsection \ref{sec_prob_bound}, the probabilistic uncertainty bound is proposed to solve the conservativeness limitation as discussed above. Consequently, the controller can gracefully degrade as the pure feedforward control if otherwise, the problem is infeasible. Moreover, the event-triggered mechanism is designed for the feedforward control to handle the probabilistic uncertainty bound and multiple external disturbances in Subsection \ref{sec_event_mecha}. Finally, the overall algorithm is summarized in Subsection \ref{sec_overall_alg}.
}

\subsection{Probabilistic Uncertainty Bound}
\label{sec_prob_bound}
To reduce the conservativeness limitation caused by the uncertainty bound $\B{W}$, the probabilistic uncertainty bound is proposed $\B{W}_{\theta}$ as
\begin{eqnarray}
P(\Delta \tilde{x}_n \in \B{W}_{\theta}) = \theta,
\end{eqnarray}
which indicates that the prediction uncertainty is bounded most of the time (with a probability $\theta$). The introduction of $\B{W}_{\theta}$ provides the flexibility to balance between robustness and conservativeness. In practice, the initial values of $\theta$ and $\B{W}_{\theta}$ can be determined by analyzing the historical information of HDVs, where numerous existing methods can be applied.  \textcolor{black}{If the problem in Eq. (\ref{pro_nom}-\ref{set_term}) is infeasible, the value of $\theta$ can be gradually decreased, such as $\theta \leftarrow \theta/2$, and, in the extreme case ($\theta=0$), the controller will gracefully degrade as the pure feedforward control.}

\textcolor{black}{
As a result of the probabilistic uncertainty bound, the mRPI set $\B{F}_{\theta}$, which is calculated based on $\B{W}_{\theta}$, is no longer a strict bound, and thus the actual tracking error $e(k)$ in Algorithm \ref{alg_tube} could exceed the planned tube without new external disturbances. In this case, the feedforward control should also be triggered to reduce the error. To handle this situation as well as the new external disturbances, an event-triggered mechanism is designed for the feedforward control in the next subsection.
}

\subsection{Event-triggered Mechanism}
\label{sec_event_mecha}
The key is to define an event $M$ as the trigger mechanism for the feedforward control. Specifically, $M$ is defined as the event that the tracking error deviation ($\tilde{e}$) exceeds the mRPI set as
\begin{eqnarray}
M =  \left\{ \tilde{e}(k-1) \in \B{F}_{\theta}, \tilde{e}(k) \notin \B{F}_{\theta}  \right\}, \nonumber
\end{eqnarray}
where $k$ indicates the current time step. \textcolor{black}{
The event $M$ could be caused either by the accumulated uncertainty or new external disturbances. When the event $M$ happens, the feedforward control of $f$-CAV is required to be triggered, and $\bar{u}_f, \bar{e}$ are replanned for the tracking error at the current time. Meanwhile, if a new external disturbance does not cause the event $M$, the disturbance can be handled as the uncertainty, and the feedforward control will not be triggered.
}

\subsection{Overall Algorithm}
\label{sec_overall_alg}
\textcolor{black}{
The framework of the robust platoon control in mixed traffic flow can be summarized in Algorithm \ref{alg_overall}. Most parts of the algorithm are the same as Algorithm \ref{alg_tube}, whereas the key differences are the introduction of the event-triggered mechanism and the probabilistic uncertainty bound. 
}

\textcolor{black}{
The overall algorithm consists of two modes. The first is the default situation without external disturbances, where no trajectory plan and prediction of $n$-HDV are available. In this mode, the controller degrades as the pure feedback control, resulting in a car-following behavior. When the event $M$ happens, the controller becomes the second mode, where the feedforward control and feedback control are integrated to control the $f$-CAV. After $N_p$ time steps, the external disturbance will be eliminated, and the controller will return to the first mode. Whenever the event $M$ happens, the feedforward control will be replanned.
}
\begin{algorithm}
	\color{black}
	\caption{Algorithm of Robust Platoon Control}
	\label{alg_overall}
	\LinesNumbered 
	Compute the feedback gain $K$\;
	Initialize $\theta$ and the probabilistic uncertainty bound $\B{W}_{\theta}$\;
	Initialize the mRPI set $\B{F}_{\theta}$ and the tight sets $\overline{\B{E}}_{\theta}, \overline{\B{U}}_{\theta}$\;
	\If{Event $M$ happens}
	{
		Compute the feedforward control $\bar{u}_f, \bar{e}$ by solving the optimization problem in Eq. (\ref{pro_nom}-\ref{set_term}), where $N_p$ is determined as discussed in Subsection \ref{sec_feedforward} and $\theta$ is determined as discussed in Subsection \ref{sec_prob_bound}\;
		\For{$k=1$ to $N_p$}
		{
			Observe the actual tracking error $e(k)$\;
			Compute the deviation $\tilde{e}(k)=e(k)-\bar{e}(k)$\;
			\If{New event $M$ happens}
			{
				Go back to the feedforward control\;
			}
			Compute the feedback control $\tilde{u}_f(k)= K \cdot  \tilde{e}(k)$\;
			Compute the actual acceleration $u_f(k) = \bar{u}_f(k) + \tilde{u}_f(k)$\;
			Implement the actual acceleration $u_f(k)$\;
		}	
	Observe the actual tracking error $e(k)$\;
	\If{New event $M$ happens}
	{
		Go back to the feedforward control\;
	}
	Compute the feedback control $\tilde{u}_f(k)= K \cdot  e(k)$\;
	Implement the feedback acceleration $\tilde{u}_f(k)$\;
	}	
	
\end{algorithm}

\section{Performance analysis}
\textcolor{black}{
	In this section, performances of the proposed framework are analyzed including feasibility and recursive feasibility, control objectives, as well as computational and communicational complexity.
}

\subsection{Feasibility and Recursive Feasibility}
\textcolor{black}{
To make Algorithm \ref{alg_overall} feasible, two conditions are required: first, the mRPI set is sufficiently small such that all tight constraints exist; second, the problem in Eq. (\ref{pro_nom}-\ref{set_term}) is feasible. For the first condition, as shown in Eq. (\ref{eq_Z}), since $A_K$ is a constant, the mRPI set can be sufficiently small, if the probabilistic uncertainty bound $\B{W}_{\theta}$ is small. In the extreme case where $\theta = 0$, $\B{W}_{\theta} = \varnothing$ and thus the mRPI sets are empty sets. Therefore, the first condition can be satisfied by adjusting the probabilistic uncertainty bound. For the second condition, the key is to reserve sufficient control time steps $N_p$ for the feedforward control to eliminate the initial tracking error and satisfy the terminal constraints, as discussed in Subsection \ref{sec_feedforward}. Please note that for $\B{W}_{\theta} = \varnothing$, Algorithm \ref{alg_overall} degrades as the MPC method (pure feedforward control), so the algorithm can have the same feasibility as the MPC method.
}

\textcolor{black}{
	Besides the feasibility, the recursive feasibility of MPC methods has also been widely investigated for a long time \cite{blanchini1999set, mayne2000constrained, cagienard2007move}. Generally, an MPC controller is called recursively feasible if and only if for all initially feasible state and for all optimal sequences of control inputs the MPC optimization problem remains feasible for all time \cite{lofberg2012oops}. In other words, if a recursively feasible MPC controller is feasible at the time step, the controller will be feasible at the next time step after conducting the optimal control input. The typical approach to achieve the recursive feasibility is to append the original MPC problem with a terminal state constraint. 
}

\textcolor{black}{
	For our algorithm, the feedforward control is only triggered by the event $M$, and thus the definition of recursive feasibility cannot be applied. Instead, we are more interested in the problem of whether the controller is feasible at the next $M$ event if it is feasible at the current $M$ event. However, it cannot be guaranteed unless additional assumptions of external disturbances are made, such as the boundedness assumption. For a large-amplitude external disturbance, if our controller is infeasible even for $\B{W}_{\theta} = \varnothing$, the MPC controller will be also infeasible at the initial state. Taking an extreme example, when an HDV recklessly cuts in front of the CAV, an extremely large external disturbance will happen, and any deceleration cannot keep the CAV satisfying the safety constraint, irrelevant to the specific controller.
}

\subsection{Control Objectives}
\textcolor{black}{For single external disturbance, }based on the tube method \cite{mayne2001robustifying}, all constraints in Eq. (\ref{set_origin}) are satisfied if the Algorithm \ref{alg_tube} is feasible. Since the safety and string stability requirements are already included in Eq. (\ref{set_origin}), safety and string stability are guaranteed. Moreover, the terminal constraints in Eq. (\ref{set_term}) are satisfied by the feedforward control. If there is no uncertainty, i.e., $\dot{v}_n=0$, the tracking error will stay at the zero value, which indicates the stability according to Definition \ref{def_individual}. \textcolor{black}{	Similarly, for multiple external disturbances, if the Algorithm \ref{alg_overall} is feasible for all $M$ events, the control objectives can also be achieved. Otherwise, as discussed above, our controller, as well as the MPC controllers, is infeasible, and the control objectives cannot be satisfied.}

\subsection{Computational and Communicational Complexity}
\textcolor{black}{
	For MPC methods, the feedforward control is triggered at each time step, where the optimization problem is solved to plan the CAV trajectory, and the planned trajectory will usually be transmitted to the upstream CAV via wireless communication. Therefore, the computational and communication complexity is determined by the frequency of feedforward control. It is also true for our framework, as the feedback control does not rely on the communication and its computational cost is negligible. Because the feedforward control is event-triggered in Algorithm \ref{alg_overall}, our method can reduce both the computational and communication burdens. Although the basic computational resources for the feedforward control should still be available, it provides huge potentials for designing more effective computational resource allocation systems \cite{zheng2015smdp, zhou2019computation}, including edge computing, in-vehicle computing, and vehicular cloud computing. Moreover, the reduced communication burden enables better wireless communication systems with less time delay and packet loss \cite{guo2016communication, wen2018cooperative, ye2019deep}.
}

\section{Numerical Experiments}

\subsection{Numerical Experiment Design}
This section conducts numerical experiments to verify the performances of the proposed control framework. First, the properties of the probabilistic uncertainty bound as well as the mRPI set are investigated. Second, experiments are designed to verify the communicational and computational efficiency, compared with the MPC method. Third, control objectives are validated regarding safety, stability, and string stability.

\begin{figure*}
	\centering
	\includegraphics[width=0.75\textwidth]{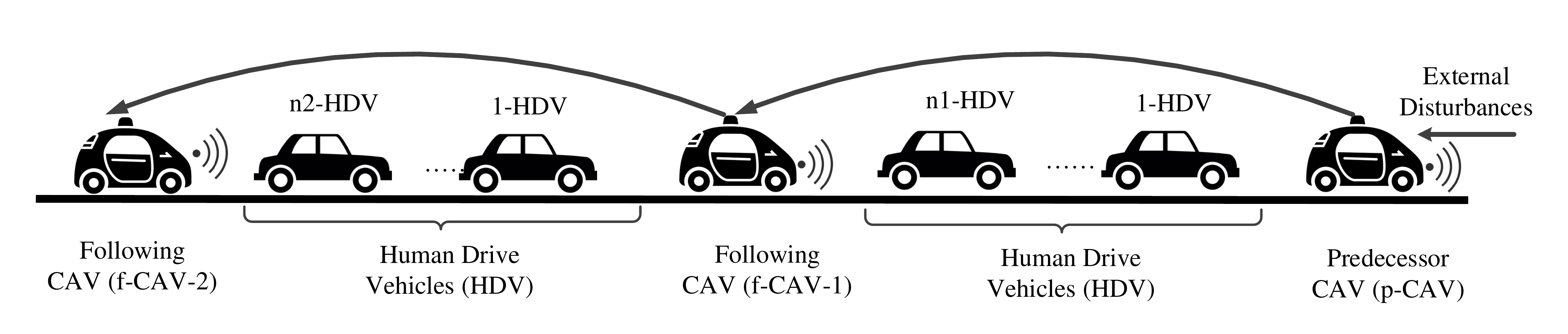}
	\caption{Illustration of the platoon in a long mixed traffic flow denoted as P-2.} 
	\label{fig_platoon_2}
\end{figure*}

To achieve the above objectives, two mixed platoons and two test scenarios are designed respectively. Besides of the platoon in Fig. \ref{fig_platoon} (denoted as P-1), a generalized platoon with another $f$-CAV is designed as shown in Fig. \ref{fig_platoon_2} (denoted as P-2). The P-2 can demonstrate the framework scalability for long mixed traffic flow. To investigate the influences of external disturbances, two test scenarios are designed involving the single external disturbance (scenario 1) and multiple external scenarios (scenario 2). A Poisson process with different $\lambda$ is utilized to generate multiple external disturbances.

As the prediction method of HDVs is not the focus of this paper, the simple car-following and prediction method is applied. The car-following behaviors of HDVs are simulated by the Newell car-following model with truncated normally distributed uncertainty and predicted by the deterministic Newell car-following model.  Specifically, the  normally distributed uncertainty is designed as $\Delta \tilde{s} \sim \C{N}(0, \sigma_s^2)$ and $\Delta \tilde{v} \sim \C{N}(0, \sigma_v^2)$, and then the distribution is truncated with the interval $[-n_s,n_s]$ and $[-n_v, n_v]$ respectively. The values of the parameters can be found in Table \ref{tab_para}.

\begin{table}
	\centering
	\footnotesize
	\setlength{\abovecaptionskip}{1pt}
	\setlength{\belowcaptionskip}{3pt}	
	\caption{The parameter values used in this paper. }
	\label{tab_para}
	\begin{tabular}{cccccc}
		\hline
		\multicolumn{1}{c}{\bfseries Parameter } &  \multicolumn{1}{c}{ \bfseries Value} &\multicolumn{1}{c}{\bfseries Parameter } &   \multicolumn{1}{c}{ \bfseries Value} &\multicolumn{1}{c}{\bfseries Parameter } &   \multicolumn{1}{c}{ \bfseries Value}  \\ \hline
		$n$  & 5 &$n_1$  & 3 &$n_2$ &3\\ 
		$\sigma_s$ &0.1 &$\sigma_v$ &0.1 &$n_s$ &1.0\\
		$n_v$ &1.0 &$\tau$ &0.5 &$h$ &0.5\\
		$v_{min}$ &0 &$v_{max}$ &50 &$u_{max}$ &5\\
		$Q$ &1 &$L$ &1 &$R$ &1\\ 
		\hline
	\end{tabular}
\end{table}

To validate the efficiency regarding computation and communication, the trigger number of the feedforward control is compared with the MPC method, which solves the problem in Eq. (\ref{pro_nom}) with $\B{Z} = \varnothing$ at each time step. To study the influence of the frequency of external disturbances, experiments are conducted in scenario 2 with different values of $\lambda$.

\subsection{Probabilistic Uncertainty Bound and mRPI Set}

\begin{figure}
	\centering
	\begin{minipage}{0.49\linewidth}
		\includegraphics[width=1\textwidth]{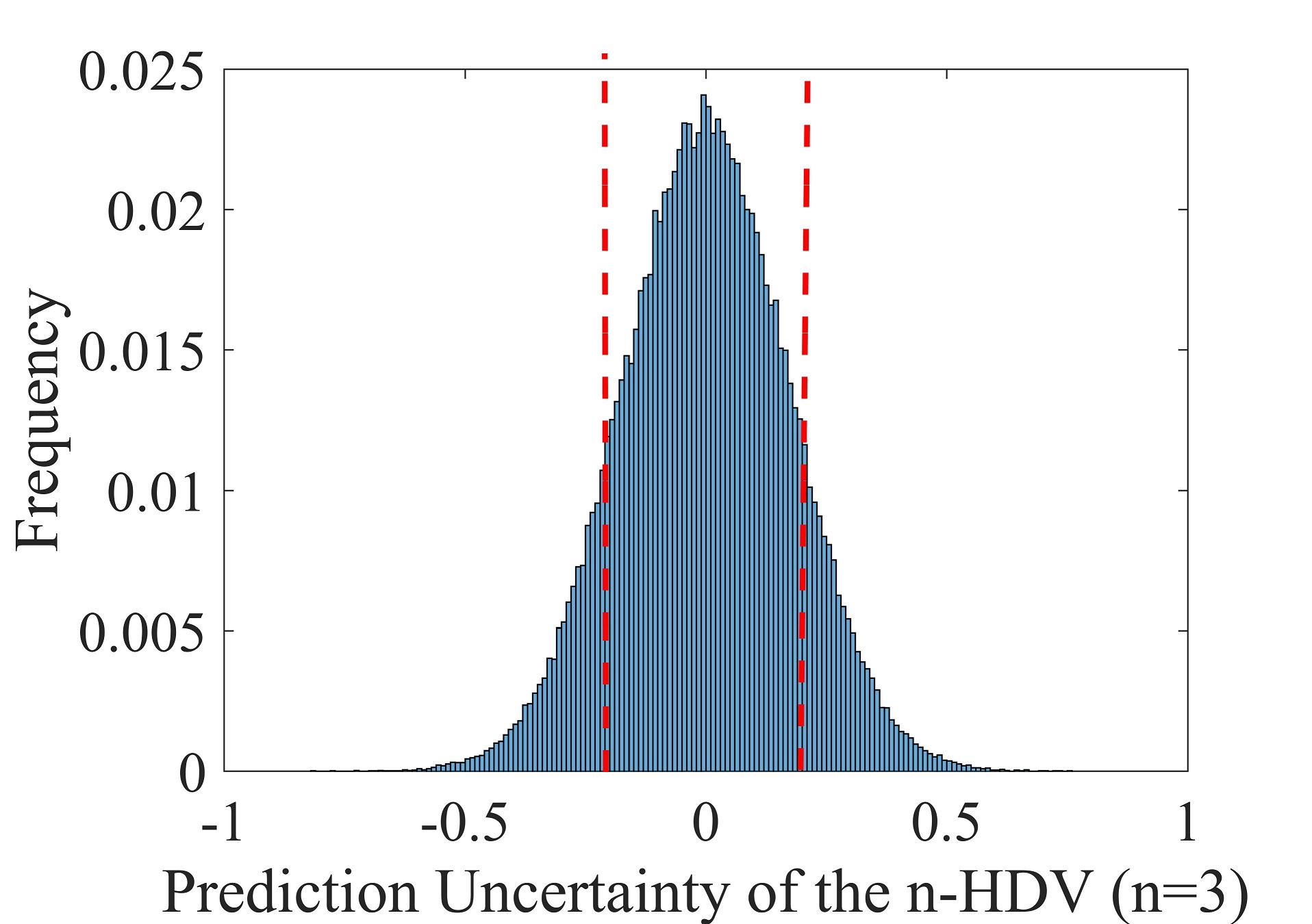}
		\centerline{(a)}
	\end{minipage}
	\begin{minipage}{0.49\linewidth}
		\includegraphics[width=1\textwidth]{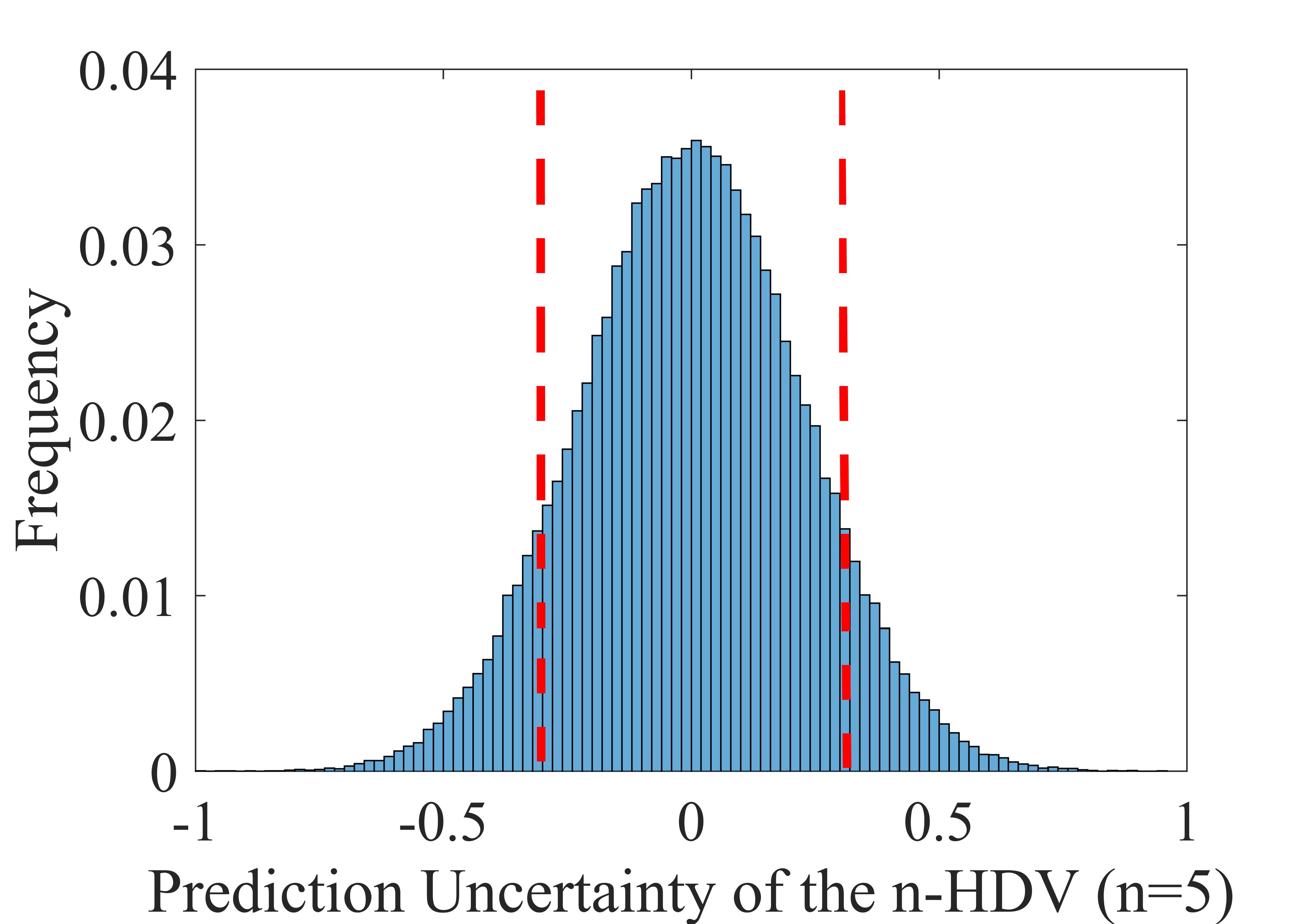}
		\centerline{(b)}
	\end{minipage}
	\begin{minipage}{0.49\linewidth}
		\includegraphics[width=1\textwidth]{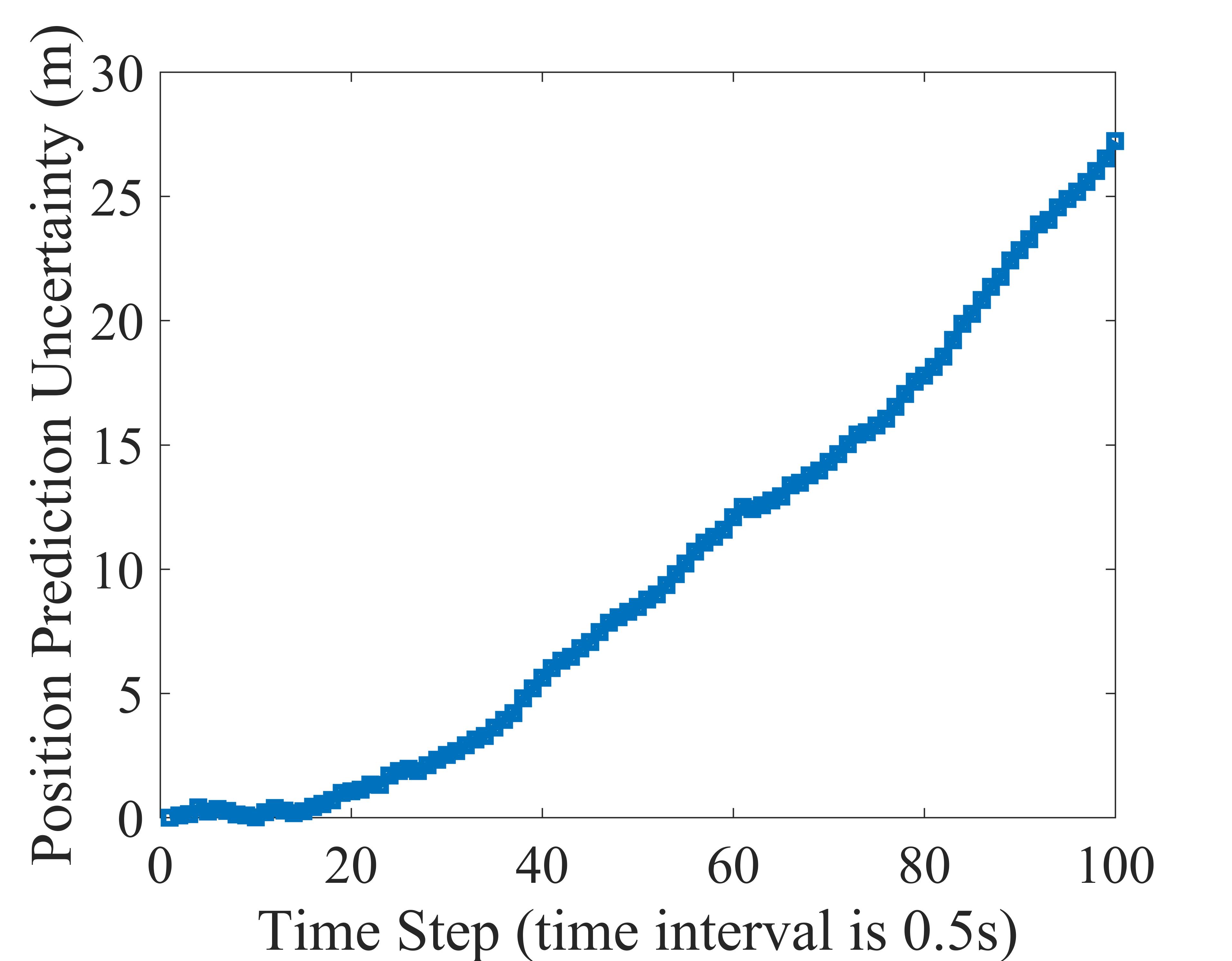}
		\centerline{(c)}
	\end{minipage}
	\begin{minipage}{0.49\linewidth}
		\includegraphics[width=1\textwidth]{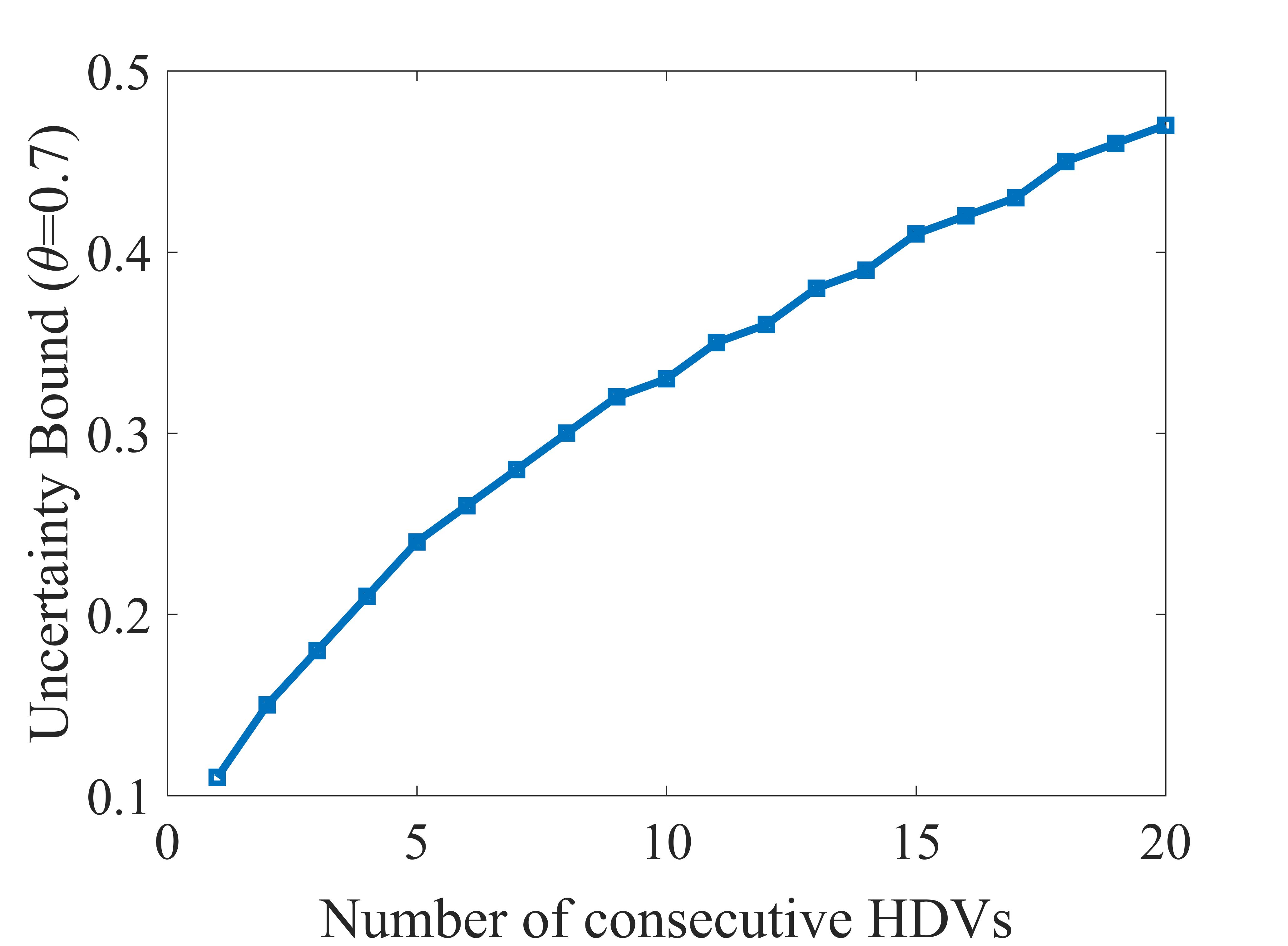}
		\centerline{(d)}
	\end{minipage}
	\begin{minipage}{0.95\linewidth}
		\includegraphics[width=1\textwidth]{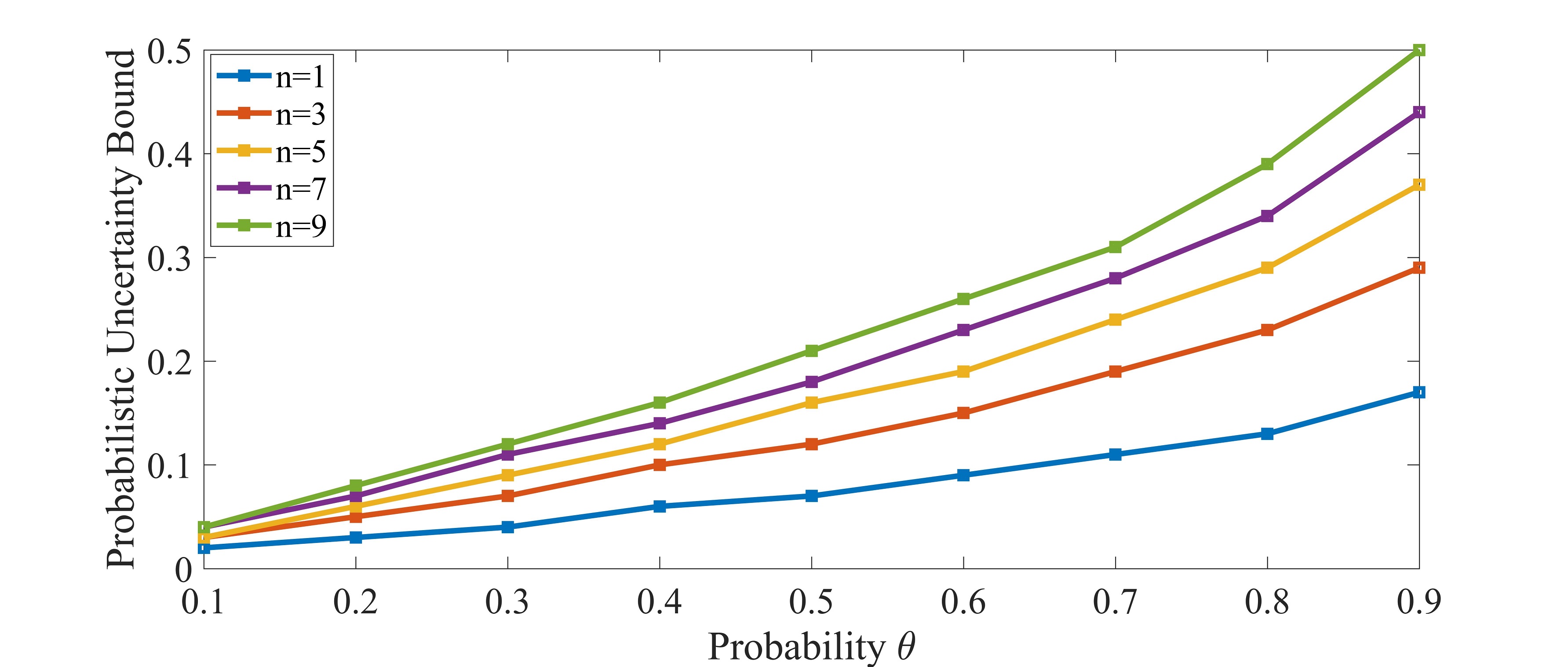}
		\centerline{(e)}
	\end{minipage}
	\caption{Simulation results of the prediction uncertainty of the $n$-HDV with $n=4$ (a), $n=5$ (b), varying time steps without feedback control (c), varying numbers of consecutive HDVs (d), and varying values of the probability $\theta$ (e).} 
	\label{fig_uncertainty}
\end{figure}

The probabilistic uncertainty bound $\B{W}_{\theta}$ is a critical parameter in our framework to balance the robustness and conservativeness. \textcolor{black}{To illustrate, two examples of the probabilistic uncertainty bound are provided in Fig. \ref{fig_uncertainty} (a, b), which shows simulation results of the $n$-HDV with $n=3$ and $n=5$ respectively.} As shown in Fig. \ref{fig_uncertainty} (a), the probabilistic uncertainty bound is set as \textcolor{black}{$\B{W}_{\theta} = \{w_s = w_v = 0.2\}$}, and the simulation results show that 
\begin{eqnarray}
P \left(\Delta \tilde{x}_3 \in \B{W}_{\theta} \right) = 0.751, \nonumber
\end{eqnarray}
\textcolor{black}{which indicates $\theta=0.751$.} Similarly, as shown in Fig. \ref{fig_uncertainty} (b), the probabilistic uncertainty bound is set as \textcolor{black}{$\B{W}_{\theta} = \{w_s = w_v = 0.3\}$} where $\theta = 0.820$ for $n=5$. 

\textcolor{black}{As pointed out in Subsection \ref{sec_challenge}, the prediction uncertainty will accumulate along the time if it is not eliminated timely, which is challenging and motivates the robust platoon control. For example, the position prediction uncertainty caused by the acceleration uncertainty will accumulate quadratically. To illustrate, the position prediction uncertainty of the $n$-HDV is simulated along the time step in Fig. \ref{fig_uncertainty} (c). Consequently, if the controller cannot mitigate the uncertainty effectively, it will lead to the system failure: constraints cannot be satisfied and control objectives cannot be achieved. Moreover, the prediction uncertainty will also be amplified spatially by consecutive HDVs. To investigate this property, we simulated different numbers of consecutive HDVs from $n=1$ to $n=20$, and obtain the uncertainty bound $\B{W}_{0.7}$. As shown in Fig. \ref{fig_uncertainty} (d), the probabilistic uncertainty bound increases linearly. Therefore, when determining the probabilistic uncertainty bound in Algorithm \ref{alg_overall}, the number of consecutive HDVs should be considered. To further investigate the influence of the probability $\theta$, we calculated the probabilistic uncertainty bounds for different number of consecutive HDVs and values of $\theta$. As shown in Fig. \ref{fig_uncertainty} (e), with the decrease of $\theta$, the probabilistic uncertainty bound $\B{W}_{\theta}$ decreases given a constant number of HDVs.}

\begin{figure}
	\centering
	\begin{minipage}{0.49\linewidth}
		\includegraphics[width=1\textwidth]{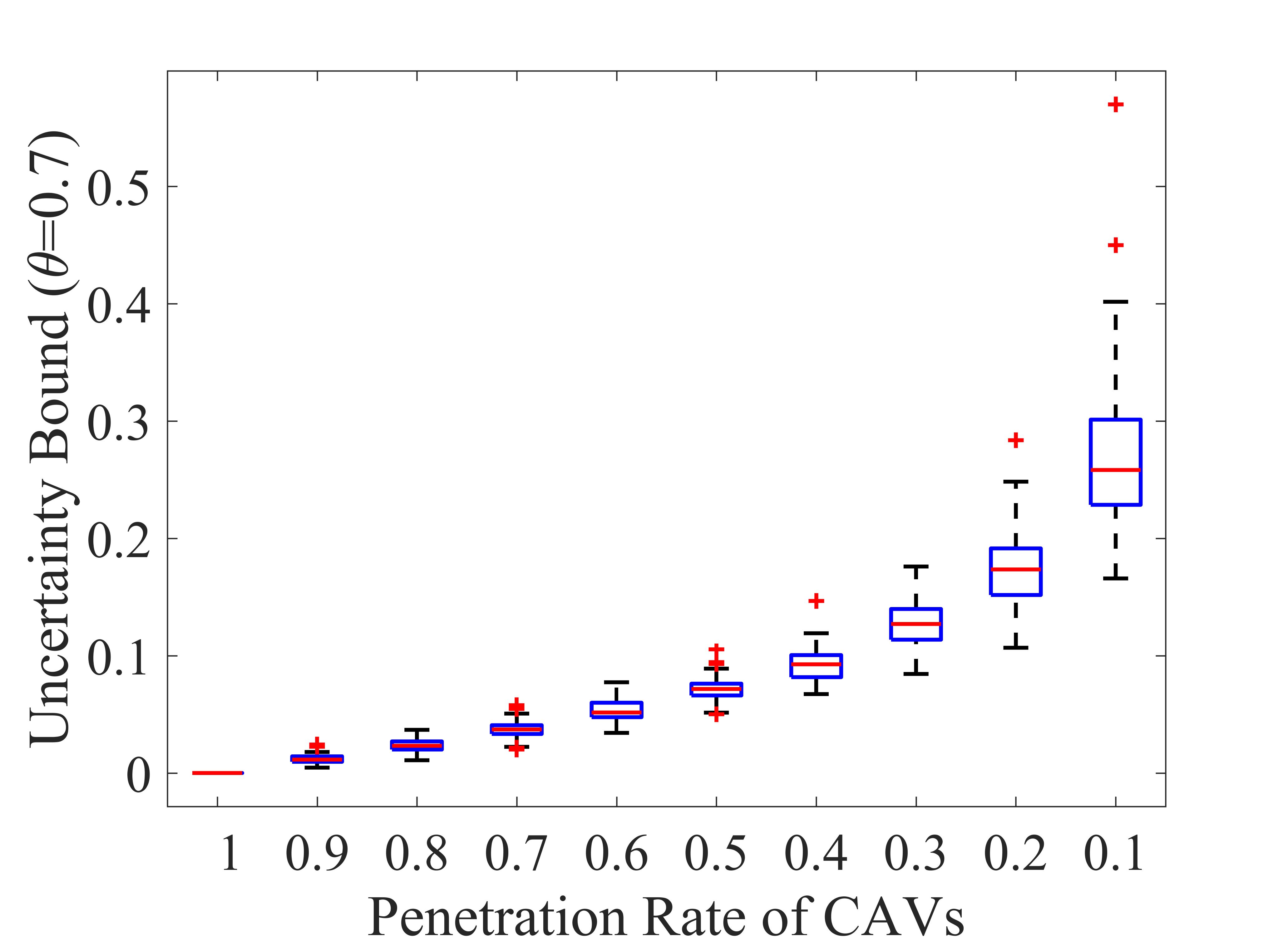}
		\centerline{(a)}
	\end{minipage}
	\begin{minipage}{0.49\linewidth}
		\includegraphics[width=1\textwidth]{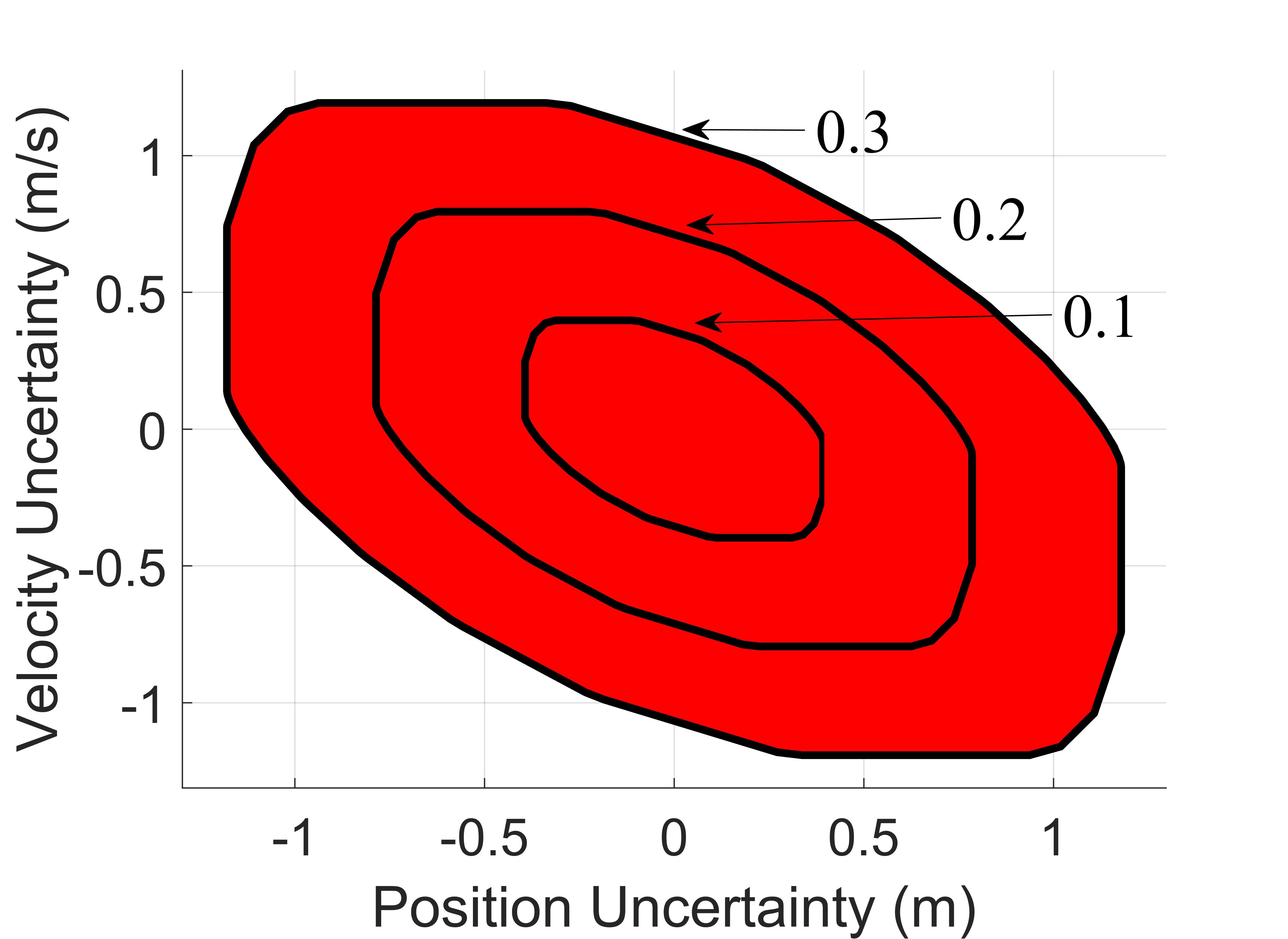}
		\centerline{(b)}
	\end{minipage}
	\caption{(a) Simulation results of the probabilistic prediction uncertainty with decreasing penetration rate of CAVs. (b) The mRPI sets for the probabilistic uncertainty bounds 0.1, 0.2, and 0.3.} 
	\label{fig_uncertainty_2}
\end{figure}

The penetration rate of the CAVs in the mixed traffic flow also affects the probabilistic uncertainty bound. To study this influence, we simulated a platoon total of 100 vehicles with different penetration rates varying from 100\% to 10\%. \textcolor{black}{For each penetration rate, the probabilistic uncertainty bounds $\B{W}_{0.7}$ for all CAVs are calculated, and the box plot is provided in Fig. \ref{fig_uncertainty_2} (a). Results show that the prediction uncertainty grows significantly with the decrease in the penetration rate. It is consistent with the intuition that CAVs can behave as stabilizers to decrease uncertainty. } 

\textcolor{black}{We further investigate the influence of probabilistic uncertainty bounds on the mRPI set. By solving Eq. (\ref{prob_feedback}), we obtain $K = [0.6406, 1.0192]$ with the parameters in Table \ref{tab_para}. Then, the mRPI set is calculated by Eq. (\ref{eq_Z}) given the probabilistic uncertainty bound. In this paper, the toolbox \cite{MPT3} is used at MATLAB 2018a with Intel i5-6200U and 16G ARM. To illustrate, three examples of mRPI sets are provided for $\B{W}=0.1, 0.2, 0.3$ in Fig. \ref{fig_uncertainty_2} (b). }

\subsection{Performance Evaluation for Platoon P-1}
\begin{figure}
	\centering
	\begin{minipage}{0.49\linewidth}
		\includegraphics[width=1\textwidth]{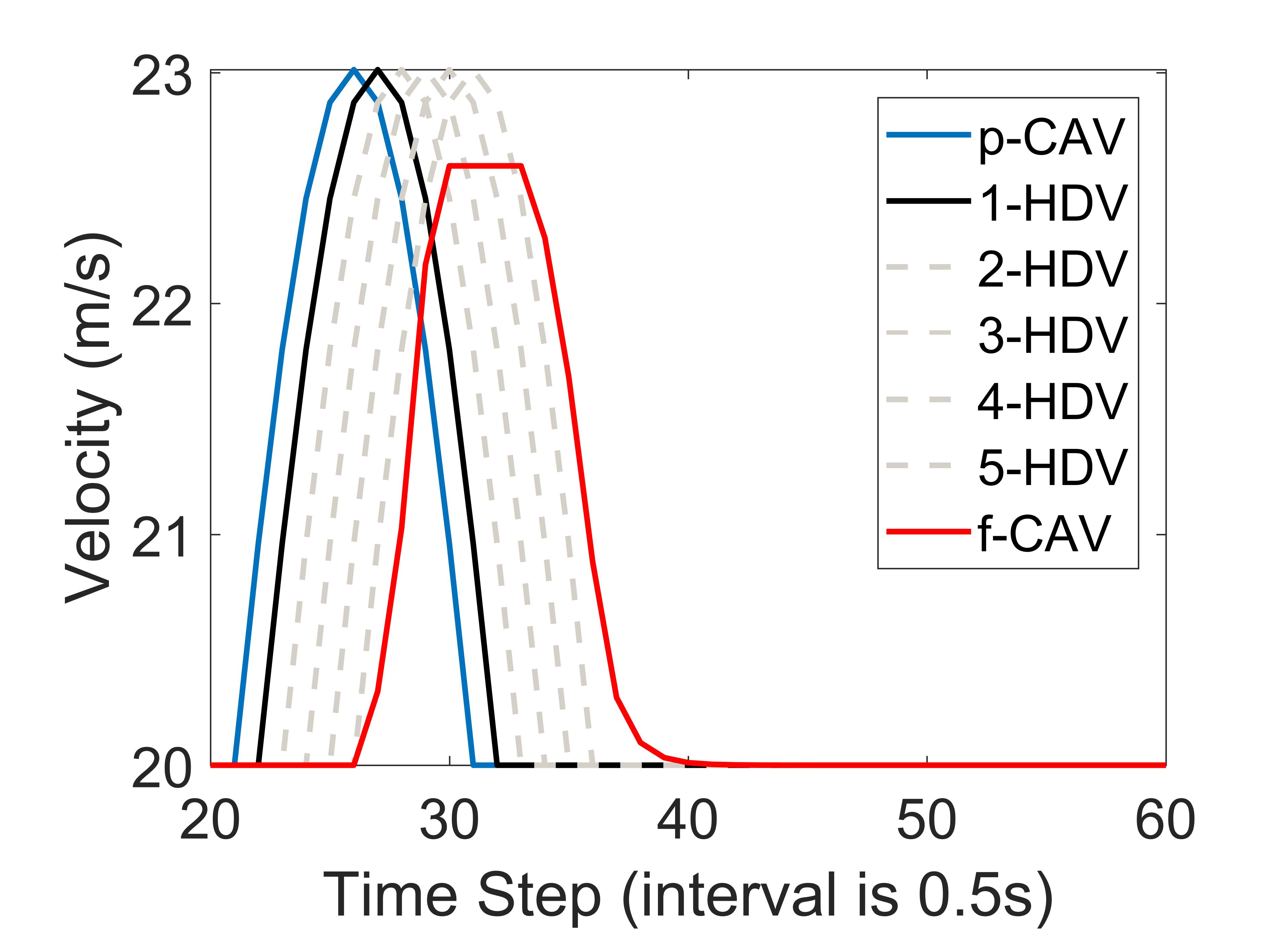}
		\centerline{(a)}
	\end{minipage}
	\begin{minipage}{0.49\linewidth}
		\includegraphics[width=1\textwidth]{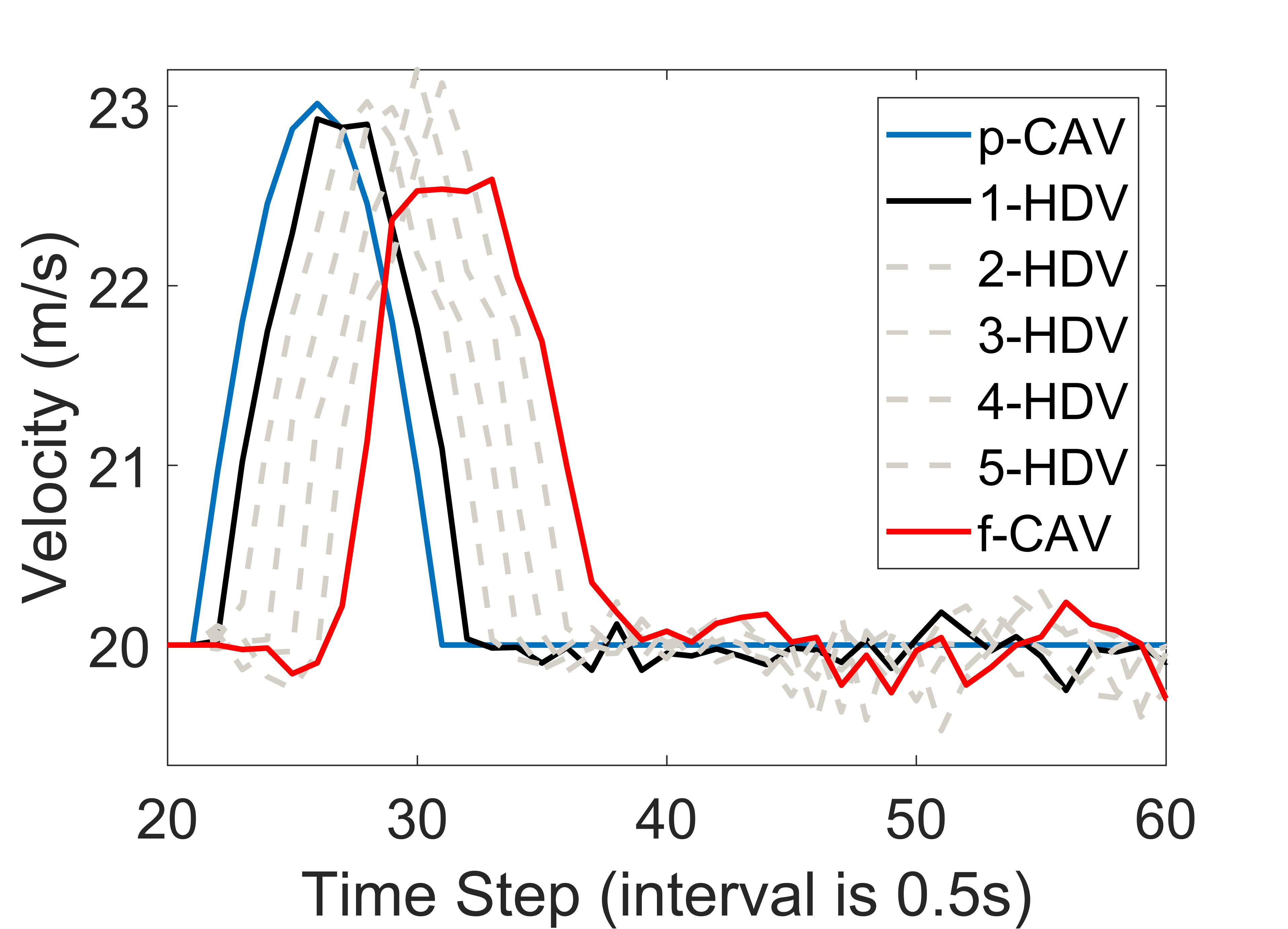}
		\centerline{(b)}
	\end{minipage}
	\begin{minipage}{0.49\linewidth}
		\includegraphics[width=1\textwidth]{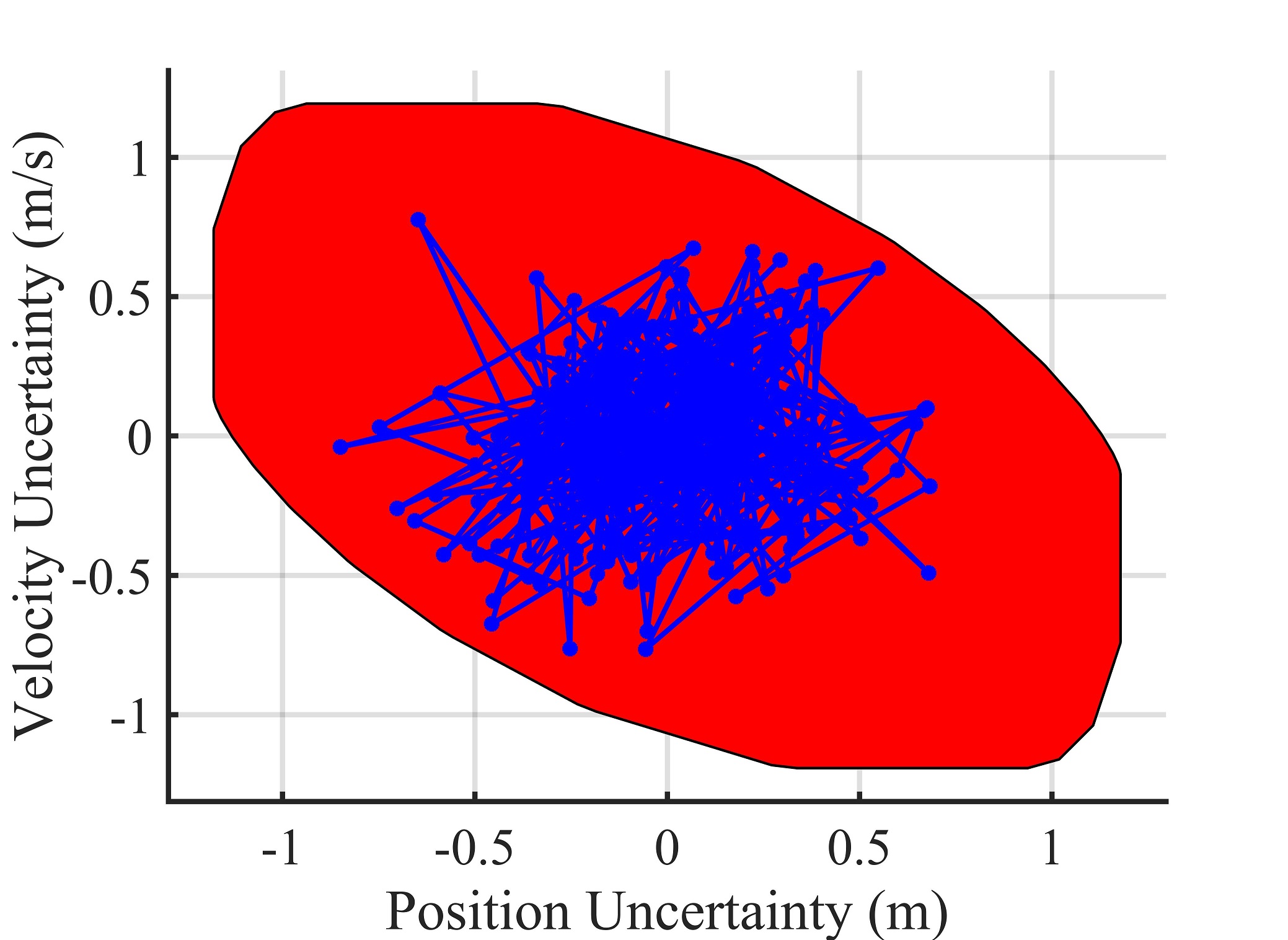}
		\centerline{(c)}
	\end{minipage}
	\begin{minipage}{0.49\linewidth}
		\includegraphics[width=1\textwidth]{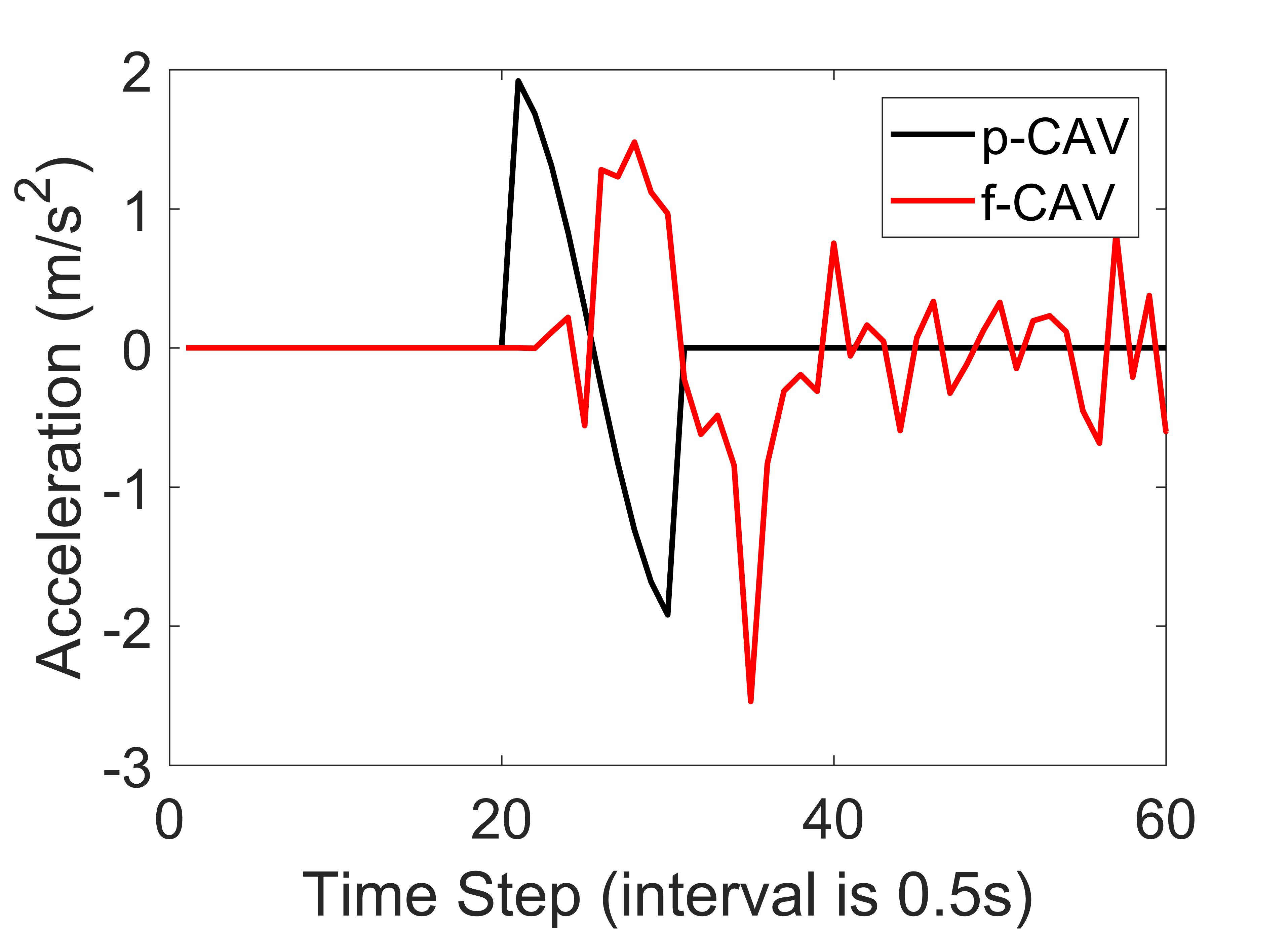}
		\centerline{(d)}
	\end{minipage}
	\caption{Simulation results of platoon P-1 in the scenario 1: without uncertainty (a); with uncertainty (b); accumulated uncertainty (c); and accelerations (d).} 
	\label{fig_results_P1_sce1}
\end{figure}

\begin{figure}
	\centering
	\begin{minipage}{0.49\linewidth}
		\includegraphics[width=1\textwidth]{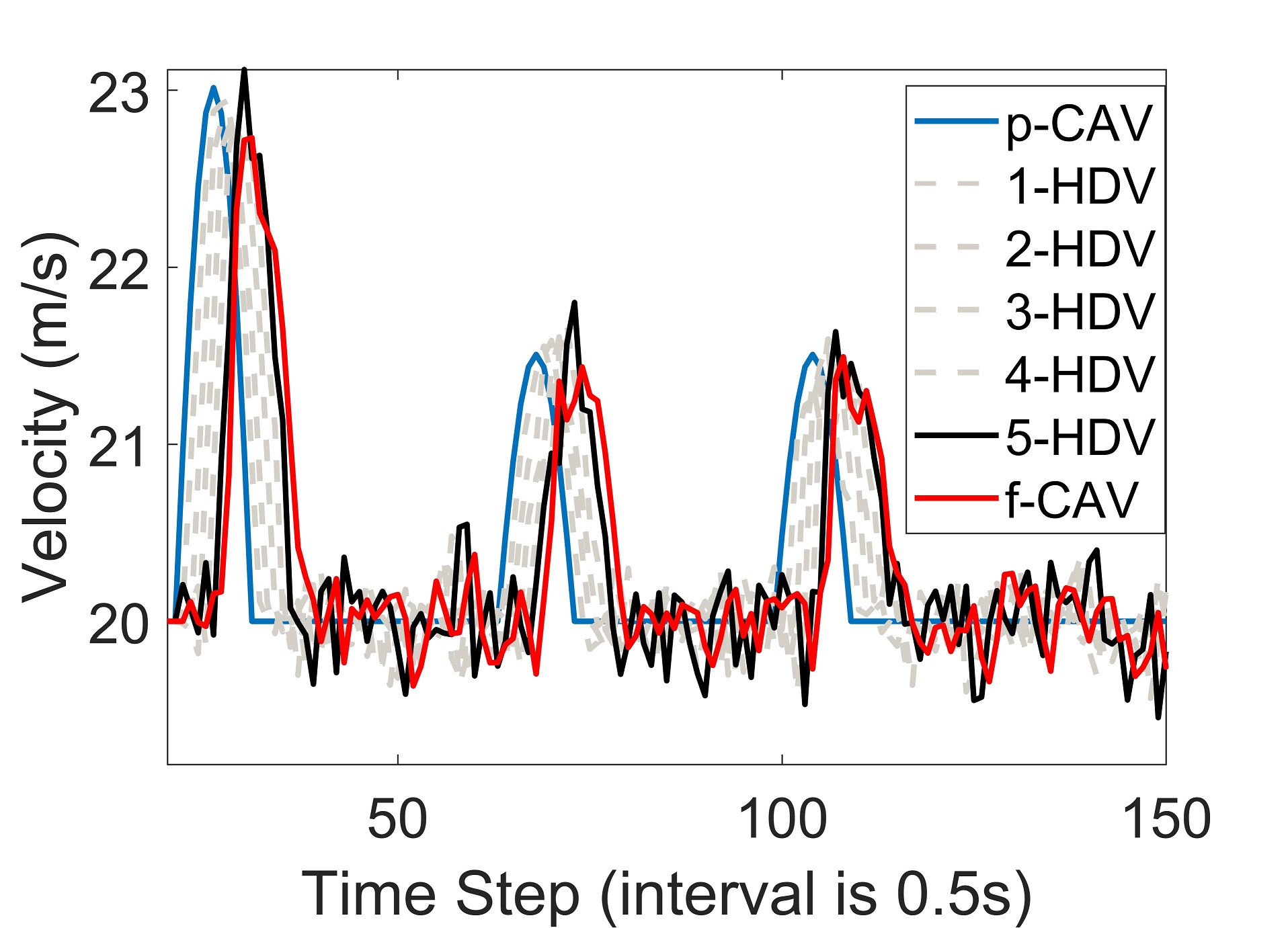}
		\centerline{(a)}
	\end{minipage}
	\begin{minipage}{0.49\linewidth}
		\includegraphics[width=1\textwidth]{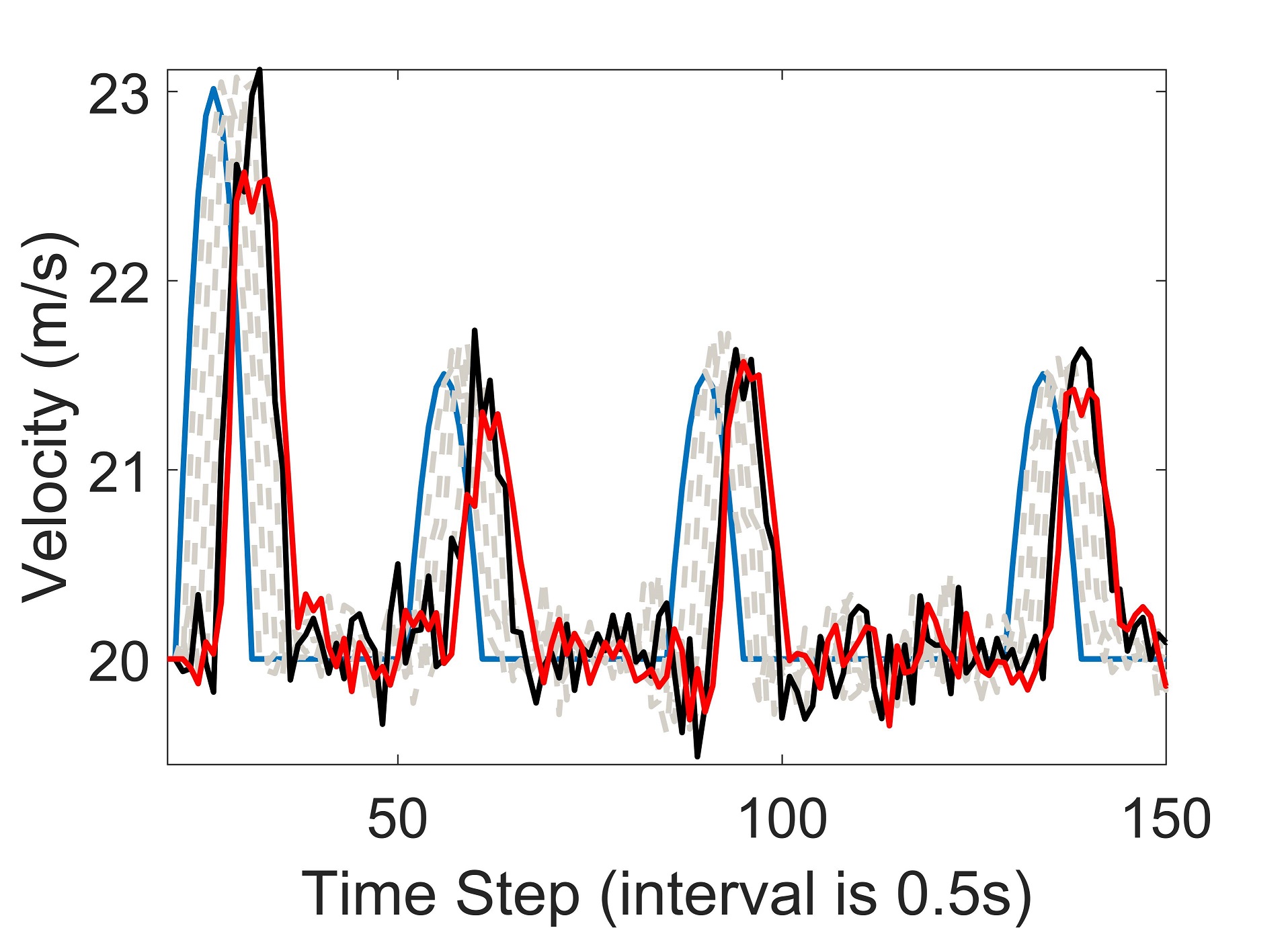}
		\centerline{(b)}
	\end{minipage}
	\begin{minipage}{0.49\linewidth}
		\includegraphics[width=1\textwidth]{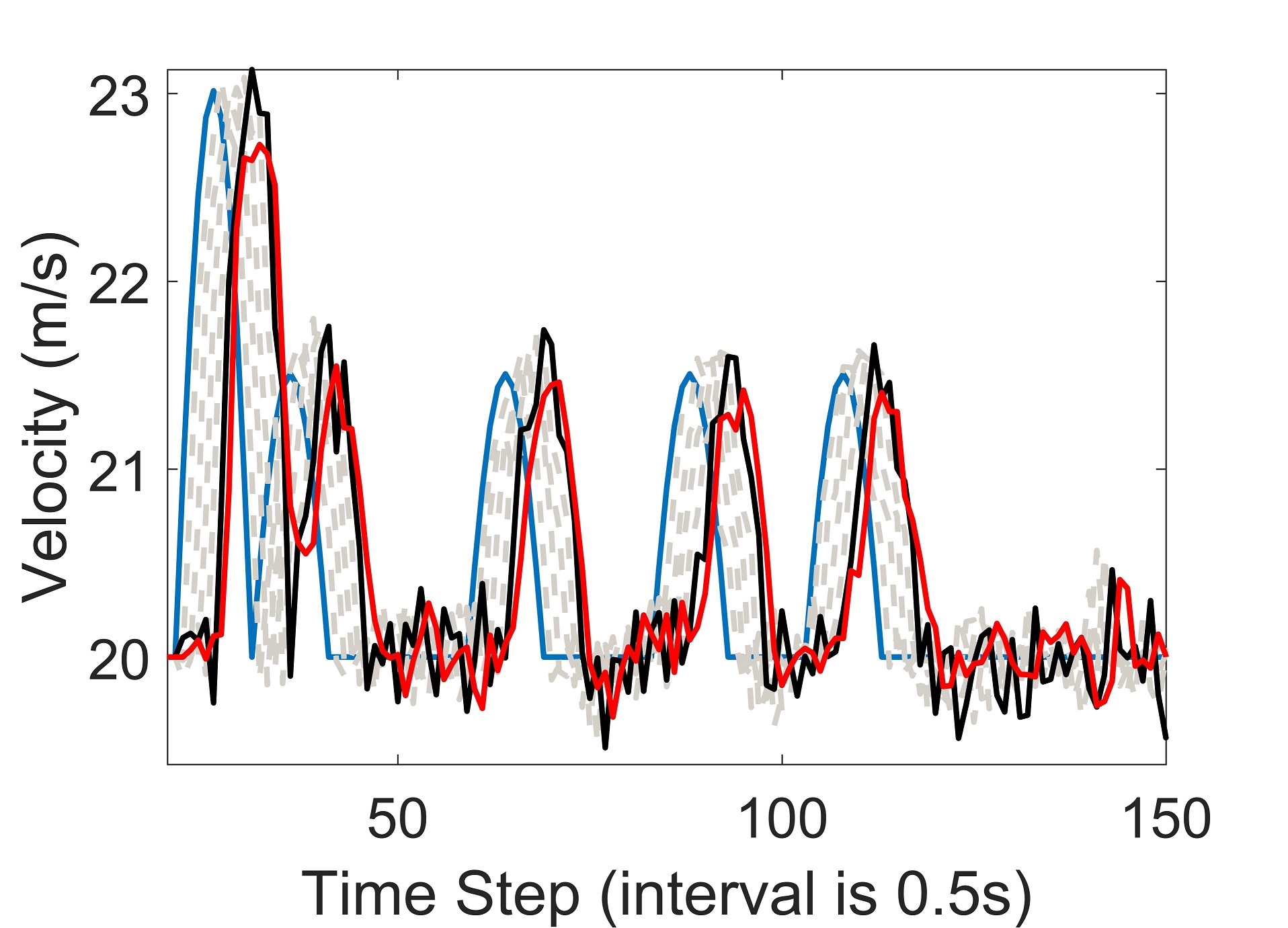}
		\centerline{(c)}
	\end{minipage}
	\begin{minipage}{0.49\linewidth}
		\includegraphics[width=1\textwidth]{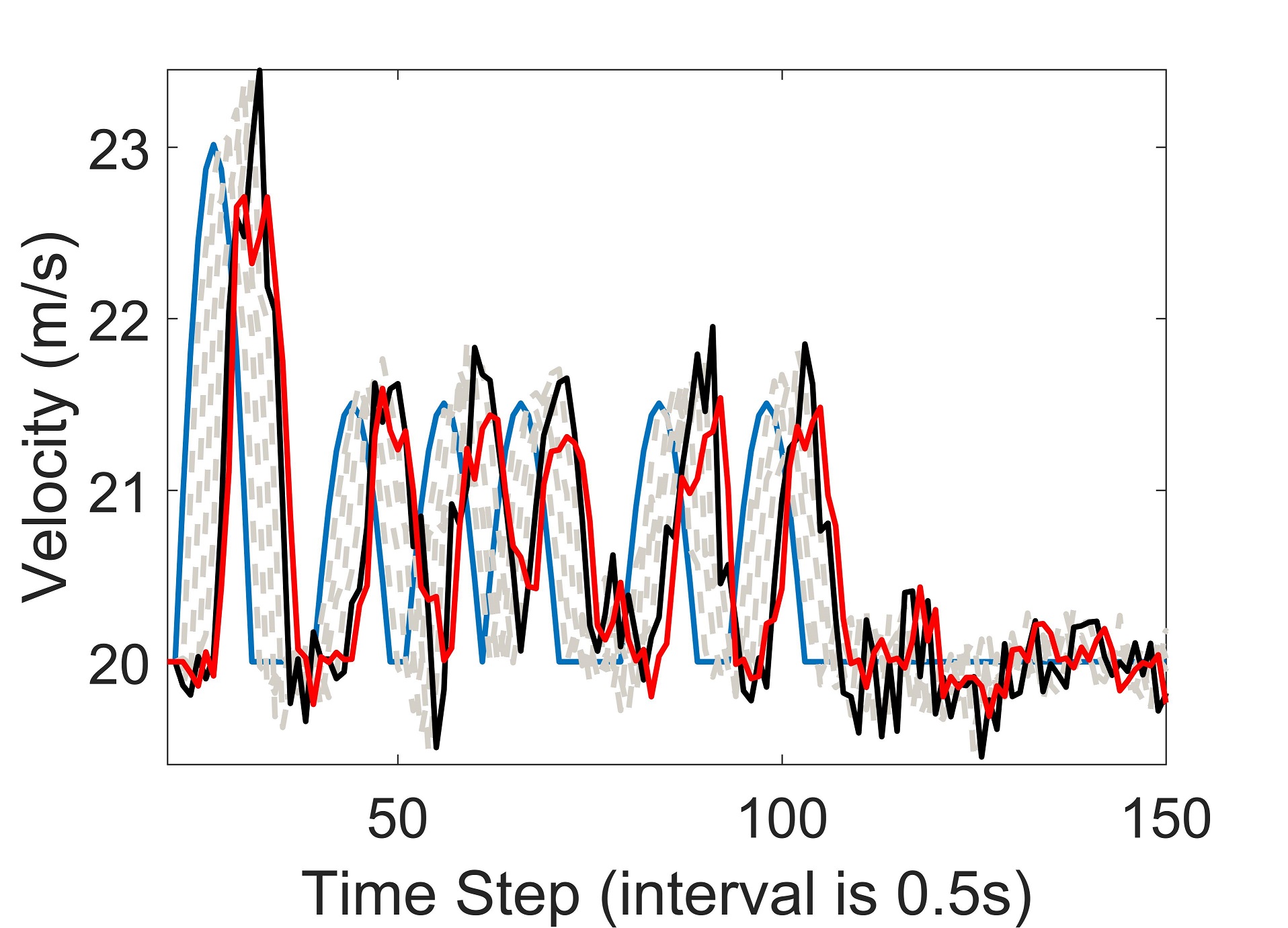}
		\centerline{(d)}
	\end{minipage}
	\caption{Simulation results of platoon P-1 in the scenario 2 with different values of $\lambda$: (a) $\lambda=10$; (b) $\lambda=7.5$; (c) $\lambda=5$; (d) $\lambda=2.5$. } 
	\label{fig_results_P1_sce2}
\end{figure}

\begin{figure}
	\centering
	\includegraphics[width=0.4\textwidth]{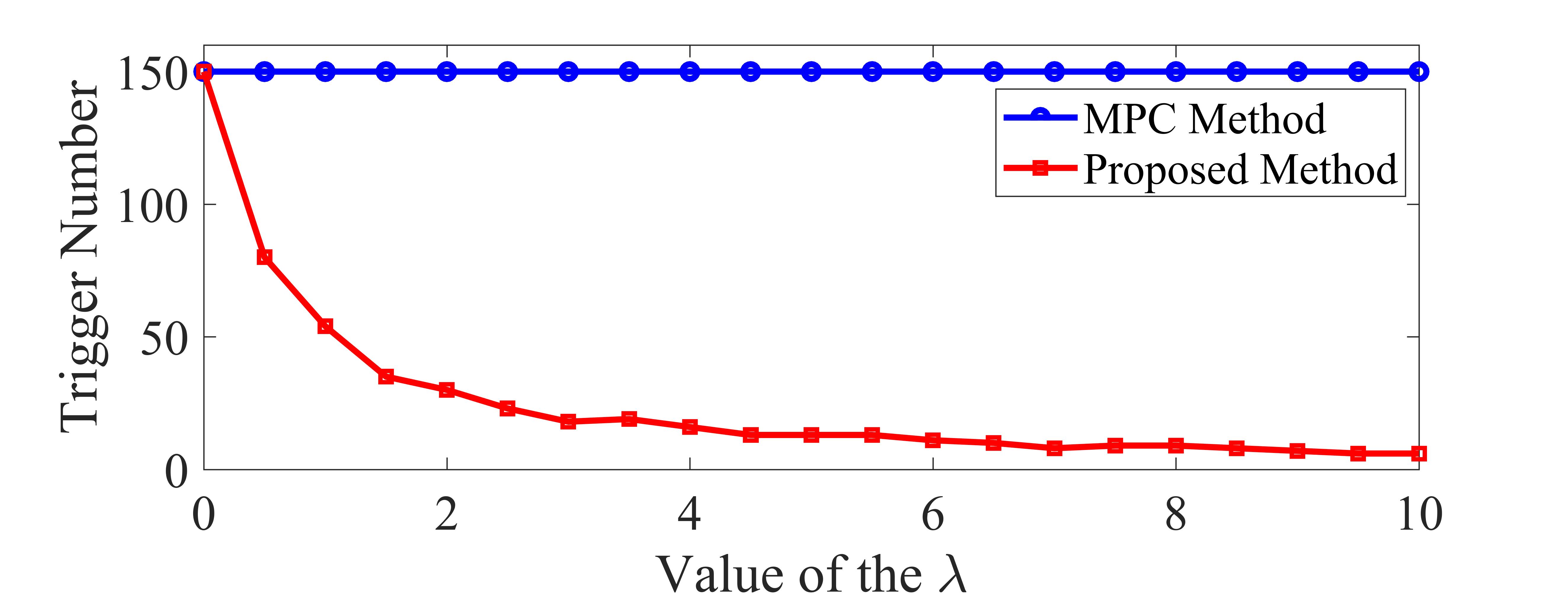}
	\caption{Trigger numbers of the feedforward control and inter-vehicular communication of platoon P-1 in scenario 2 with different values of $\lambda$. } 
	\label{fig_triggerNum_P1}
\end{figure}

This subsection evaluates the performance of the proposed method in the platoon P-1. Fig. \ref{fig_results_P1_sce1} shows the results at scenario 1, where only initial external disturbance exists. \textcolor{black}{To better interpret the proposed method, the feedforward control and feedback control are studied, respectively.} Fig. \ref{fig_results_P1_sce2} shows the results in scenario 2, where multiple external disturbances emerge following the Poisson progress with different values of $\lambda$. \textcolor{black}{To validate the computational and communicational efficiency, the trigger numbers of the proposed method and the MPC method are compared as shown in Fig. \ref{fig_triggerNum_P1}.}

Fig. \ref{fig_results_P1_sce1} (a) illustrates the planned trajectory of the $f$-CAV to handle the initial external speed disturbance, if there is no prediction uncertainty of HDVs. \textcolor{black}{Specifically, all HDVs follow their predecessors, while the $f$-CAV optimizes its trajectory to eliminate the speed disturbance for converging to the equilibrium speed (20 m/s). Fig. \ref{fig_results_P1_sce1} (b) illustrates the actual trajectory of the $f$-CAV considering the prediction uncertainty of HDVs. As the CAV is controlled by both the feedforward and feedback controllers, the actual trajectory fluctuates around the planned trajectory in Fig. \ref{fig_results_P1_sce1} (a). After the $N_p=50$ time steps, the actual trajectory converges to a small zone around the equilibrium speed, where only the feedback control is utilized to mitigate the uncertainty. Meanwhile, the maximal speed disturbance is decreased by the $f$-CAV compared with the $p$-CAV, though it is increased by HDVs. By Definition \ref{def_individual} and \ref{def_lp}, the CAVs are stable and the mixed platoon is string stable.}

To validate the effectiveness of mRPI set, all the tracking error disturbances of the $f$-CAV are analyzed. As shown in Fig. \ref{fig_results_P1_sce1} (c), all the disturbances are bounded by the mRPI set. \textcolor{black}{It guarantees that the actual trajectory is always inside the planned tube, so all constraints are satisfied. To further interpret the proposed controller,} Fig. \ref{fig_results_P1_sce1} (d) demonstrates the actual acceleration of the $f$-CAV, which is the combination of the planned acceleration $\bar{u}$ and the acceleration of the feedback control $\tilde{u}$. \textcolor{black}{The major trend of the $f$-CAV's acceleration, i.e., first accelerate and then decelerate, is determined by the feedforward control, while the fluctuations are determined by the feedback control.}

Fig. \ref{fig_results_P1_sce2} further demonstrates the ability of the proposed method in handling multiple external disturbances (scenario 2). \textcolor{black}{To investigate the influence of the disturbance frequency, different values of $\lambda$ are utilized to generate the external disturbances, namely $\lambda=10, 7.5, 5, 2.5$. The black line denotes the trajectory of the $p$-CAV, the black line denotes the trajectory of the $5$-HDV, and the red line denotes the trajectory of the $f$-CAV. As shown in Fig. \ref{fig_results_P1_sce2}, with the decrease of the $\lambda$, the external disturbance happens more frequently. For all the four cases, the $f$-CAV's velocity can eventually converge to the zone around the equilibrium velocity if no new external disturbance happens. Moreover, the maximal disturbance of the $f$-CAV is smaller than that of the $p$-CAV, though it can be amplified by the $5$-HDV. Therefore, for multiple external disturbances with different frequencies, the proposed controller can also guarantee stability and string stability.}

Fig. \ref{fig_triggerNum_P1} compares the trigger numbers of the feedforward control and inter-vehicular communication between the MPC method and the proposed method with different values of $\lambda$. Since the MPC method solves the optimization problem at each control interval, which relies on the communication information, the trigger number of the MPC method keeps equal to the total simulation steps (i.e., 150). For the proposed method, the feedforward control, as well as the inter-vehicular communication, is only triggered when the $M$ event happens. Therefore, with the increasing of the $\lambda$, the occurring number of the external disturbances decreases, so the trigger number of feedforward control and communication decreases. Considering the computational burden mainly comes from the feedforward control, the proposed method is efficient regarding computation and inter-vehicular communication. \textcolor{black}{Although the basic computational resources for the feedforward control should still be available, it provides huge potentials for designing more effective in-vehicle computational resource allocation systems \cite{zheng2015smdp, zhou2019computation}. Moreover, the reduced communication burden enables better wireless communication systems with less time delay and packet loss \cite{guo2016communication, wen2018cooperative, ye2019deep}.}

\subsection{Performance Evaluation for Platoon P-2}
\begin{figure}
	\centering
	\begin{minipage}{0.49\linewidth}
		\includegraphics[width=1\textwidth]{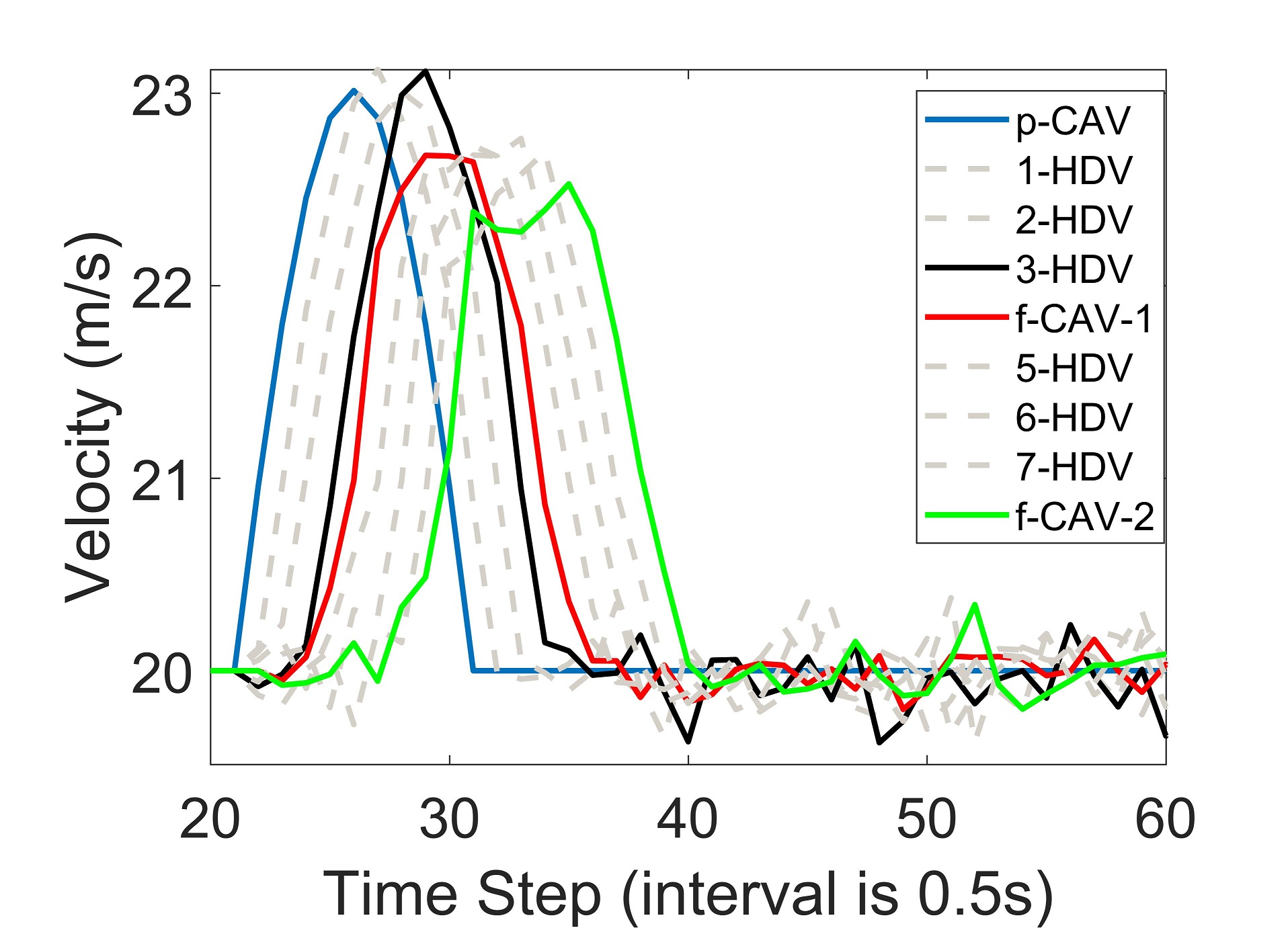}
		\centerline{(a)}
	\end{minipage}
	\begin{minipage}{0.49\linewidth}
		\includegraphics[width=1\textwidth]{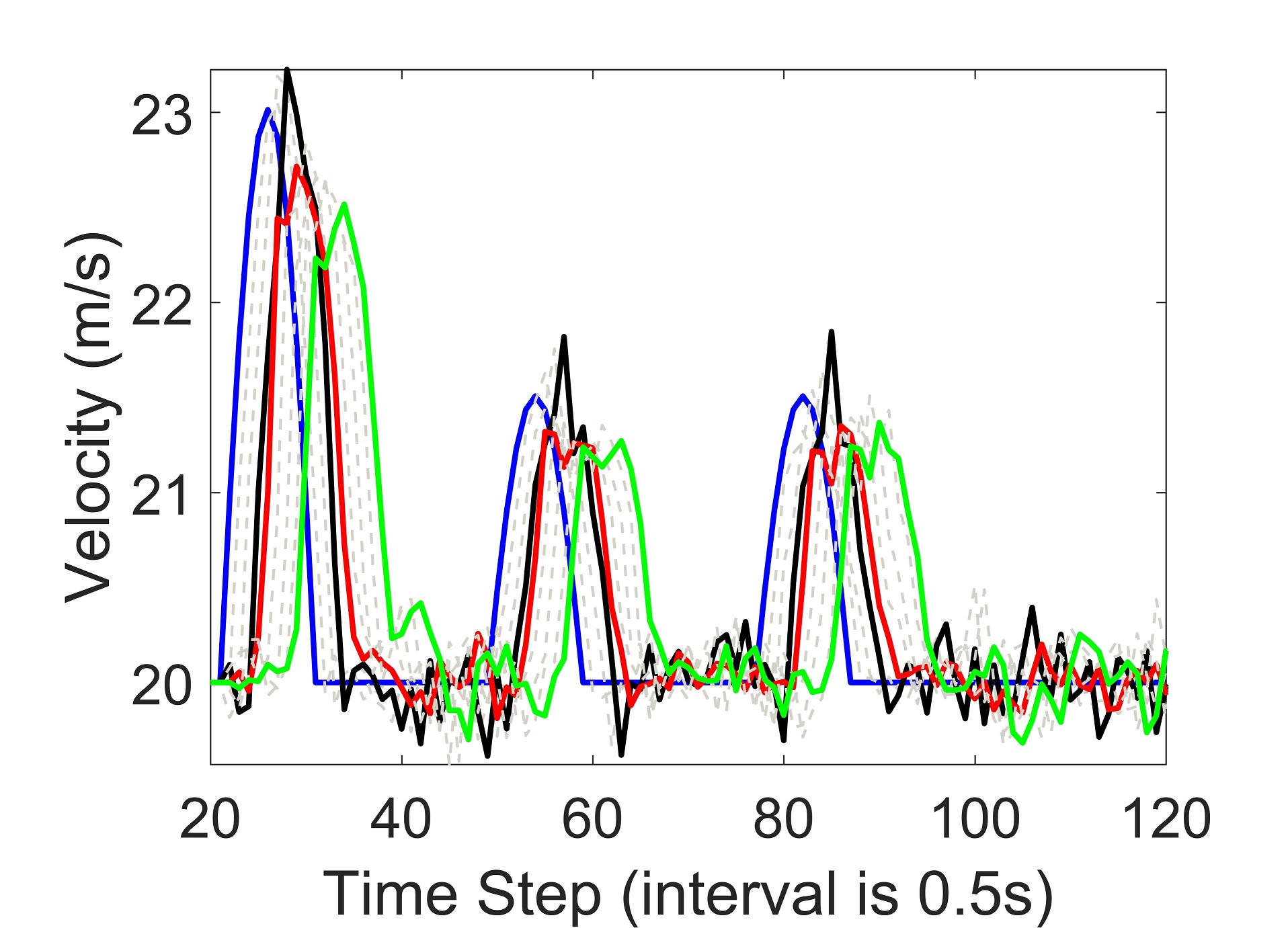}
		\centerline{(b)}
	\end{minipage}
	\begin{minipage}{0.49\linewidth}
		\includegraphics[width=1\textwidth]{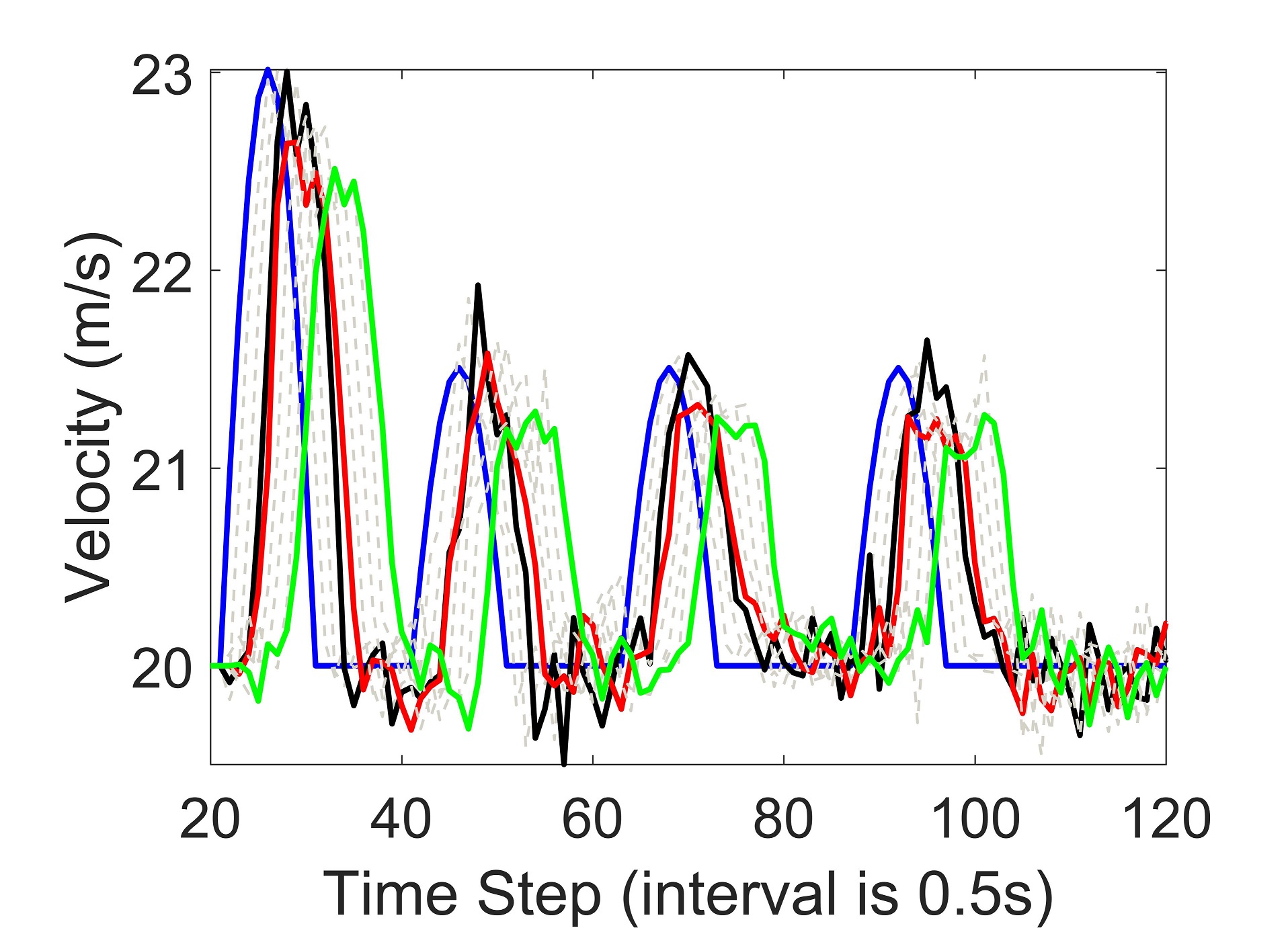}
		\centerline{(c)}
	\end{minipage}
	\begin{minipage}{0.49\linewidth}
		\includegraphics[width=1\textwidth]{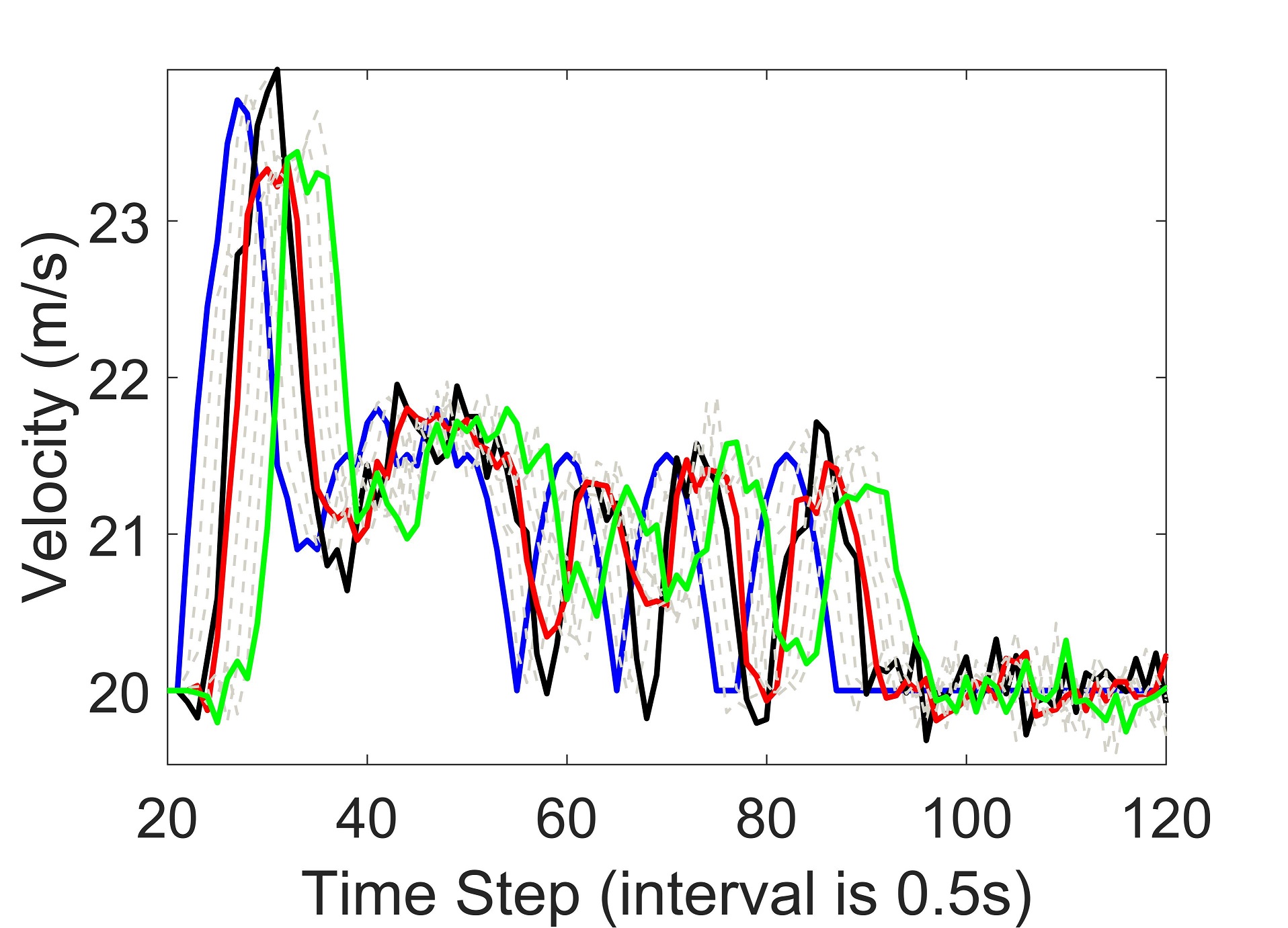}
		\centerline{(d)}
	\end{minipage}
	\caption{Simulation results of the platoon P-2 in different scenarios: (a) scenario 1; (b) scenario 2 with $\lambda=7.5$; (c) scenario 2 with $\lambda=5$; (d) scenario 2 with $\lambda=2.5$.} 
	\label{fig_result_P2}
\end{figure}

This subsection further evaluates the scalability of the proposed framework in the long mixed traffic flow. \textcolor{black}{The proposed framework is implemented at the platoon P-2 for both scenario 1 and 2.  Fig. \ref{fig_result_P2} shows the simulation results with different external disturbances. The black line denotes the $p$-CAV's velocity, the black line denotes the $3$-HDV's velocity, the red line denotes the $f$-CAV-1's velocity, and the green line denotes the $f$-CAV-2's velocity.} 

\textcolor{black}{Fig. \ref{fig_result_P2} (a) shows the results for single external disturbance. Compared with Fig. \ref{fig_results_P1_sce1} (b), the $f$-CAV-1's trajectory of P-2 is very similar to the $f$-CAV's trajectory of P-1. It is reasonable because the $f$-CAV-1's situation at P-2 is similar to the $f$-CAV at P-1. After the feedforward control of the $f$-CAV-1 is triggered, its planned trajectory is transmitted to the $f$-CAV-2, and then the feedforward control of the $f$-CAV-2 is also triggered. From the perspective of $f$-CAV-2, the $f$-CAV-1 is the predecessor and plays the role of $p$-CAV. Results show both the CAVs can stabilize the velocity and decrease the maximal disturbance, namely stable and string stable.} 

Fig. \ref{fig_result_P2} (b-d) further investigate the controller performance with multiple external disturbances. \textcolor{black}{Similar to Fig. \ref{fig_results_P1_sce2}, different values of $\lambda$ are utilized to generate the external disturbances, namely $\lambda=7.5, 5, 2.5$. Compared with Fig. \ref{fig_results_P1_sce2}, the $f$-CAV-1's trajectory has similar performances as the $f$-CAV. For the $f$-CAV-2, the controller can also effectively mitigate the disturbances even for the consecutive external disturbances as shown in Fig. \ref{fig_result_P2} (d). All these results validate the ability of the proposed framework for long mixed traffic flow.}

\section{Conclusions}
\label{sec_Con}
In this paper, a robust platoon control framework is proposed for mixed traffic flow based on tube MPC. The challenge of prediction uncertainty of HDVs is addressed \textcolor{black}{by considering a probabilistic uncertainty bound. The feedback control is designed to restrict the uncertainty inside the minimal robust positively invariant set for most of the time, while, for other time, the feedforward control is triggered based on tube MPC. The event $M$ is designed for this event-triggered mechanism. Compared with state-of-the-art platoon control methods based on MPC, our framework can handle the uncertainty explicitly with less burden of communication and computation. By adjusting the probabilistic uncertainty bound, the framework has the flexibility to balance between the feedforward control and feedback control. The performances of the proposed method are validated by numerical experiments. }

There are several promising future studies following this research. First, the integration of advanced prediction models of HDVs can further improve the control performance. Second, taking the speed guidance from intersection control as external disturbance, the proposed framework is promising for the joint control of signal and mixed traffic flow.

\bibliographystyle{IEEEtran}
\bibliography{TRC-07152020}

\begin{IEEEbiography}[{\includegraphics[width=1in,height=1.25in,clip,keepaspectratio]{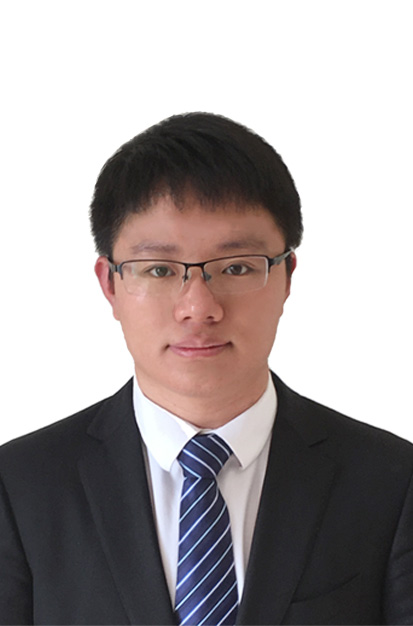}}]{Shuo Feng}
	received the bachelor`s and Ph.D. degrees in department of automation, Tsinghua University, China, in 2014 and 2019, respectively. He was also a joint Ph.D. student in civil and environmental engineering with University of Michigan, Ann Arbor, MI, USA, from 2017 to 2019, where he is currently a Post-Doctoral Researcher.	His current research interests include connected and automated vehicle evaluation, mixed traffic control, and transportation data analysis.
\end{IEEEbiography}

\begin{IEEEbiography}[{\includegraphics[width=0.9in,height=1.2in,clip,keepaspectratio]{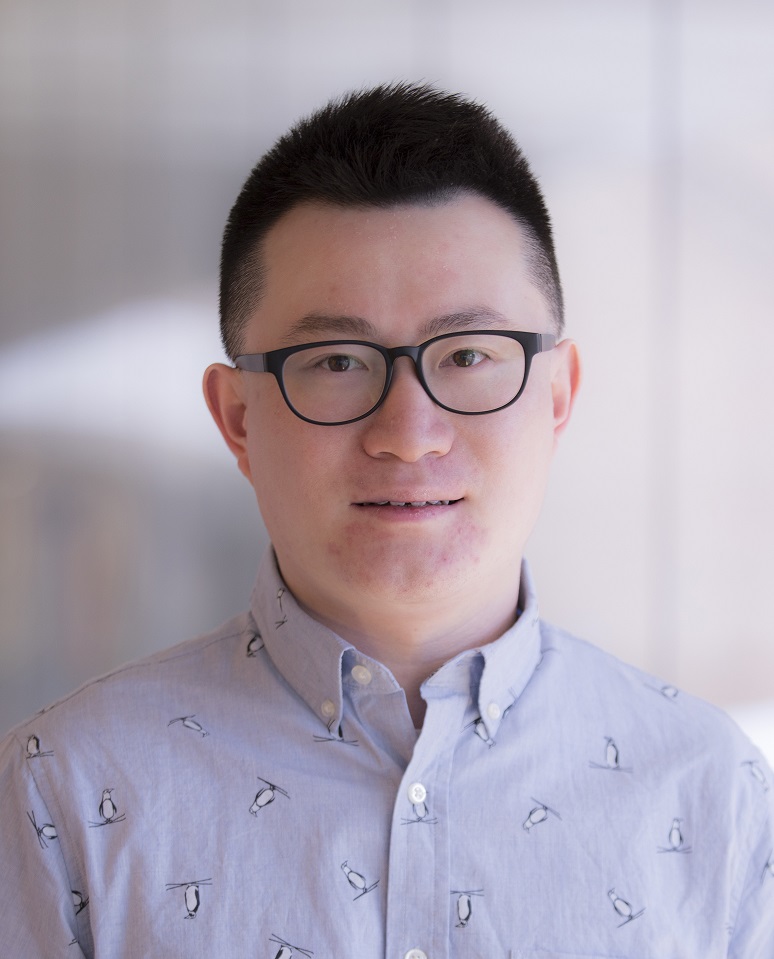}}]{Ziyou Song}
	 received his B.E. degree (with honor) and Ph.D. degree (with highest honor and thesis award) in Automotive Engineering from Tsinghua University, Beijing, China, in 2011 and 2016, respectively. He worked as a Research Scientist at Tsinghua University from 2016 to 2017 and a Research Fellow at the University of Michigan from 2017-2019. He is currently an Assistant Research Scientist/Lecturer in the Department of Electrical Engineering and Computer Science, at the University of Michigan, Ann Arbor. His research interests lie in the areas of modeling, estimation, optimization, and control of energy storage for electrified vehicles and renewable systems. Dr. Song is the author or co-author of more than 50 peer-reviewed publications including 40 journal articles. He has received several paper awards, including Applied Energy 2015-2016 Highly Cited Paper Award, Applied Energy Award for Most Cited Energy Article from China, NSK Outstanding Paper Award of Mechanical Engineering, and 2013 IEEE VPPC Best Student Paper Award.
\end{IEEEbiography}

\begin{IEEEbiography}[{\includegraphics[width=1in,height=1.25in,clip,keepaspectratio]{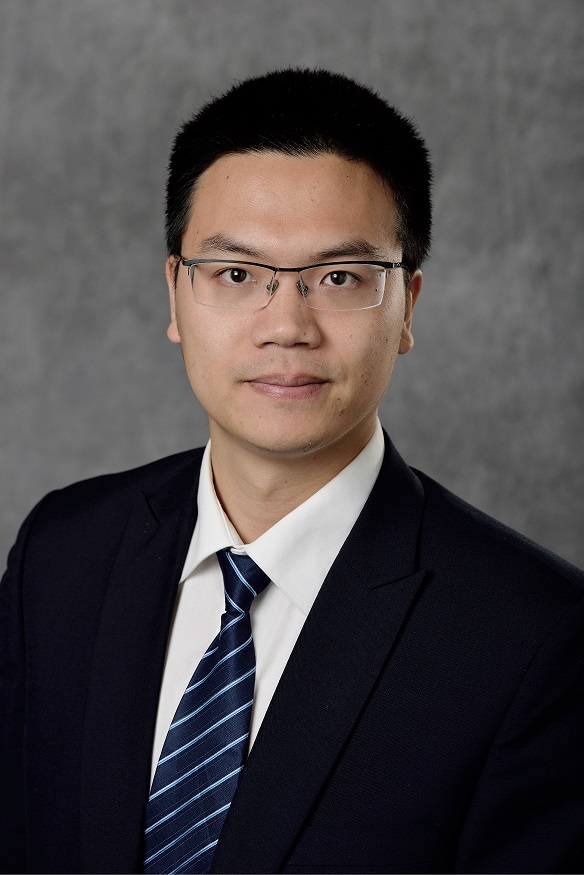}}]{Zhaojian Li}
	received his B. Eng. degree from Nanjing University of Aeronautics and Astronautics in 2010. He obtained M.S. (2013) and Ph.D. (2015) in Aerospace Engineering (flight dynamics and control) at the University of Michigan, Ann Arbor. Dr. Li worked as an algorithm engineer at General Motors from January 2016 to July 2017. Since August 2017, he has been an Assistant Professor in the department of Mechanical Engineering at Michigan State University. His research interests include Learning-based Control, Nonlinear and Complex Systems, and Robotics and Automated Vehicles.
\end{IEEEbiography}

\begin{IEEEbiography}[{\includegraphics[width=1in,height=1.25in,clip,keepaspectratio]{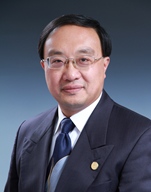}}]{Yi  Zhang}
	received   the   BS   degree   in1986   and   MS   degree   in   1988  from Tsinghua University in China, and earned  the  Ph.D.  degree  in  1995  from  the University  of  Strathclyde  in  UK.  He  is a   professor   in   the   control   science   and engineering  at  Tsinghua  University  with his  current  research  interests  focusing  on intelligent  transportation  systems. His  active  research  areas include  intelligent  vehicle-infrastructure  cooperative  systems, analysis  of  urban  transportation  systems,  urban  road  network management,  traffic  data  fusion  and  dissemination,  and  urban traffic control and management.   His research fields also cover the advanced control theory and applications, advanced detection and measurement, systems engineering, etc.
\end{IEEEbiography}

\begin{IEEEbiography}[{\includegraphics[width=1in,height=1.25in,clip,keepaspectratio]{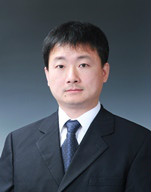}}]{Li Li}
	is currently an associate professor with Department of Automation, Tsinghua University, Beijing, China, working in the fields of complex and networked systems, intelligent control and sensing, intelligent transportation systems,  and intelligent  vehicles.  He  received  the Ph.D. degree in systems and industrial engineering from University of Arizona, USA, in 2005.  Dr. Li had published over 70 SCI indexed international journal papers and over 70 international conference papers as a first/corresponding author. 
\end{IEEEbiography}

\end{document}